\documentclass[twocolumn]{aastex63}

\usepackage{amsfonts,amsmath,graphicx,natbib,subfigure,color,gensymb,longtable,footnote}

\def\Msol{\rm{M_{\odot}}}

\def\sn{CSS161010}
\def\cow{AT\,2018cow}
\def\koala{ZTF18abvkwla}

\received{March 25, 2020}
\revised{ -- }
\accepted{April 23, 2020}
\submitjournal{ApJL}

\shorttitle{The mildly relativistic fast-rising blue optical transient CSS161010}
\shortauthors{Coppejans et al. }

\graphicspath{{./}{figures/}}

\begin{document}

\title{A mildly relativistic outflow from the energetic, fast-rising blue optical transient CSS161010 in a dwarf galaxy}

\correspondingauthor{Deanne Coppejans}
\email{deanne.coppejans@northwestern.edu}

\author[0000-0001-5126-6237]{Coppejans D.L.}
\affiliation{Center for Interdisciplinary Exploration and Research in Astrophysics (CIERA) and Department of Physics and Astronomy, Northwestern University, Evanston, IL 60208}

\author[0000-0003-4768-7586]{Margutti R.}
\altaffiliation{Alfred P. Sloan Fellow.}
\affiliation{Center for Interdisciplinary Exploration and Research in Astrophysics (CIERA) and Department of Physics and Astronomy, Northwestern University, Evanston, IL 60208}
\affiliation{CIFAR Azrieli Global Scholars program, CIFAR, Toronto, Canada}

\author[0000-0003-0794-5982]{Terreran G.}
\affiliation{Center for Interdisciplinary Exploration and Research in Astrophysics (CIERA) and Department of Physics and Astronomy, Northwestern University, Evanston, IL 60208}

\author[0000-0002-8070-5400]{Nayana A.J.}
\affiliation{Department of physics, United Arab Emirates University, Al-Ain, UAE, 15551}
\affiliation{National Centre for Radio Astrophysics, Tata Institute of Fundamental Research, PO Box 3, Pune, 411007, India}

\author[0000-0003-3765-6401]{Coughlin E.R.}
\affiliation{Department of Astrophysical Sciences, Peyton Hall, Princeton University, Princeton, NJ 08544, USA}

\author[0000-0003-1792-2338]{Laskar T.}
\affiliation{Department of Physics, University of Bath, Claverton Down, Bath, BA2 7AY, United Kingdom}

\author[0000-0002-8297-2473]{Alexander K.D.}
\altaffiliation{Einstein Fellow}
\affiliation{Center for Interdisciplinary Exploration and Research in Astrophysics (CIERA) and Department of Physics and Astronomy, Northwestern University, Evanston, IL 60208}

\author{Bietenholz M.}
\affiliation{Hartebeesthoek Radio Observatory, P.O. Box 443, Krugersdorp, 1740, South Africa}
\affiliation{Department of Physics and Astronomy, York University, Toronto, M3J 1P3, Ontario, Canada}

\author[0000-0003-0939-8775]{Caprioli D.}
\affiliation{Department of Astronomy and Astrophysics, University of Chicago, Chicago, IL 60637, USA}

\author[0000-0002-0844-6563]{Chandra P.}
\affiliation{National Centre for Radio Astrophysics, Tata Institute of Fundamental Research, PO Box 3, Pune, 411007, India}

\author[0000-0001-7081-0082]{Drout M. R.}
\affiliation{Department of Astronomy and Astrophysics, University of Toronto, 50 St. George Street, Toronto, Ontario, M5S 3H4 Canada}
\affiliation{The Observatories of the Carnegie Institution for Science, 813 Santa Barbara St., Pasadena, CA 91101, USA 31}

\author{Frederiks D.}
\affiliation{Ioffe Institute, Polytekhnicheskaya, 26, St. Petersburg, 194021, Russian Federation}

\author[0000-0001-9553-4723]{Frohmaier C.}
\affiliation{Institute of Cosmology and Gravitation, University of Portsmouth, Portsmouth, PO1 3FX, UK}

\author[0000-0003-3315-1975]{Hurley K.H}
\affiliation{University of California at Berkeley, Space Sciences Laboratory, 7 Gauss Way, Berkeley, CA 94720, USA}

\author[0000-0001-6017-2961]{Kochanek C.S.}
\affiliation{Center for Cosmology and AstroParticle Physics (CCAPP), The Ohio State University, 191 W. Woodruff Avenue, Columbus, OH 43210, USA}
\affiliation{Department of Astronomy Ohio State University, 140 W. 18th Ave., Columbus, OH 43210, USA}

\author[0000-0002-1417-8024]{MacLeod M.}
\affiliation{Center for Astrophysics | Harvard and Smithsonian, 60 Garden Street, Cambridge, MA, 02138, USA}

\author[0000-0002-1125-7384]{Meisner A.}
\affiliation{NSF's National Optical-Infrared Astronomy Research Laboratory,  950 N Cherry Ave, Tucson, AZ 85719, USA}

\author[0000-0002-3389-0586]{Nugent P.E.}
\affiliation{Lawrence Berkeley National Laboratory, 1 Cyclotron Road, Berkeley, CA 94720, USA}

\author{Ridnaia A.}
\affiliation{Ioffe Institute, Polytekhnicheskaya, 26, St. Petersburg, 194021, Russian Federation}

\author[0000-0003-4102-380X]{Sand D. J.}
\affiliation{Steward Observatory, University of Arizona, 933 North Cherry Avenue, Tucson, AZ 85721-0065, USA}

\author{Svinkin D.}
\affiliation{Ioffe Institute, Polytekhnicheskaya, 26, St. Petersburg, 194021, Russian Federation}

\author{Ward C.}
\affiliation{Department of Astronomy, University of Maryland, College Park, MD 20742, USA}
\affiliation{Lawrence Berkeley National Laboratory, 1 Cyclotron Road, Berkeley, CA 94720, USA}

\author{Yang S.}
\affiliation{Department of Physics, University of California, 1 Shields Avenue, Davis, CA 95616-5270, USA}
\affiliation{INAF Osservatorio Astronomico di Padova, Vicolo dell’Osservatorio 5, I-35122 Padova, Italy}
\affiliation{The Oskar Klein Centre, Department of Astronomy, Stockholm University, AlbaNova, SE-106 91 Stockholm, Sweden}

\author{Baldeschi A.}
\affiliation{Center for Interdisciplinary Exploration and Research in Astrophysics (CIERA) and Department of Physics and Astronomy, Northwestern University, Evanston, IL 60208}

\author[0000-0002-7924-3253]{Chilingarian I.V.}
\affiliation{Center for Astrophysics | Harvard and Smithsonian, 60 Garden Street, Cambridge, MA, 02138, USA}
\affiliation{Sternberg Astronomical Institute, M.~V.~Lomonosov Moscow State University, 13 Universitetsky prospect, Moscow, 119234, Russia}

\author[0000-0002-9363-8606]{Dong Y.}
\affiliation{Department of Physics and Astronomy, Purdue University, 525 Northwestern Avenue, West Lafayette, IN 47907, USA}

\author{Esquivia C.}
\affiliation{Hamilton College, 198 College Hill Road, Clinton, NY 13323}
\affiliation{Center for Interdisciplinary Exploration and Research in Astrophysics (CIERA) and Department of Physics and Astronomy, Northwestern University, Evanston, IL 60208}

\author[0000-0002-7374-935X]{Fong W.}
\affiliation{Center for Interdisciplinary Exploration and Research in Astrophysics (CIERA) and Department of Physics and Astronomy, Northwestern University, Evanston, IL 60208}

\author[0000-0001-6869-0835]{Guidorzi C.}
\affiliation{Department of Physics and Earth Science, University of Ferrara, via Saragat 1, I-44122, Ferrara, Italy}

\author{Lundqvist, P.}
\affiliation{Department of Astronomy, AlbaNova University Center, Stockholm University, SE-10691 Stockholm, Sweden}
\affiliation{The Oskar Klein Centre, AlbaNova, SE-10691 Stockholm, Sweden.}

\author[0000-0002-0763-3885]{Milisavljevic D.}
\affiliation{Department of Physics and Astronomy, Purdue University, 525 Northwestern Avenue, West Lafayette, IN 47907, USA}

\author[0000-0001-8340-3486]{Paterson K.}
\affiliation{Center for Interdisciplinary Exploration and Research in Astrophysics (CIERA) and Department of Physics and Astronomy, Northwestern University, Evanston, IL 60208}

\author[0000-0002-5060-3673]{Reichart D.E.}
\affiliation{Department of Physics and Astronomy, University of North Carolina at Chapel Hill, Chapel Hill, NC 27599, USA}

\author[0000-0003-4631-1149]{Shappee B.}
\affiliation{Institute for Astronomy, University of Hawai'i, 2680 Woodlawn Drive, Honolulu, HI 96822, USA}

\author[0000-0002-3019-4577]{Stroh M.C.}
\affiliation{Center for Interdisciplinary Exploration and Research in Astrophysics (CIERA) and Department of Physics and Astronomy, Northwestern University, Evanston, IL 60208}

\author[0000-0001-8818-0795]{Valenti S.}
\affiliation{Department of Physics, University of California, 1 Shields Avenue, Davis, CA 95616-5270, USA}

\author{Zauderer B.A.}
\affiliation{National Science Foundation, 2415 Eisenhower Ave., Alexandria, VA 22314, USA}
\affiliation{Neils Bohr Institute, University of Copenhagen, Blegdamsvej 17, 2100 Copenhagen, Denmark}

\author[0000-0002-9725-2524]{Zhang B.}
\affiliation{Department of Physics and Astronomy, University of Nevada, Las Vegas, NV 89154, USA}



\begin{abstract}
We present X-ray and radio observations of the Fast Blue Optical Transient (FBOT) CRTS-CSS161010 J045834-081803 
(\sn{} hereafter) at $t=69-531$ days. \sn{} shows luminous X-ray ($L_x\sim5\times 10^{39}\,\rm{erg\,s^{-1}}$) and radio ($L_{\nu}\sim10^{29}\,\rm{erg\,s^{-1}Hz^{-1}}$) emission. The radio emission peaked at $\sim$100 days post transient explosion and rapidly decayed. We interpret these observations in the context of synchrotron emission from an expanding blastwave. \sn{} launched a mildly relativistic outflow with velocity $\Gamma\beta c\ge0.55c$ at $\sim100$ days. This is faster than the non-relativistic \cow{} ($\Gamma\beta c\sim0.1c$) and closer to \koala{} ($\Gamma\beta c\ge0.3c$ at 63 days). The inferred initial kinetic energy of \sn{} ($E_k\gtrsim10^{51}$ erg) is comparable to that of long Gamma Ray Bursts (GRBs), but the ejecta mass that is coupled to the mildly relativistic outflow is significantly larger ($\sim0.01-0.1\,\Msol$). This is consistent with the lack of observed $\gamma$-rays. The luminous X-rays were produced by a different emission component to the synchrotron radio emission. \sn{} is located at $\sim$150 Mpc in a dwarf galaxy with stellar mass $M_{*}\sim10^{7}\,\rm{M_{\odot}}$ and specific star formation rate $sSFR\sim 0.3\,\rm{Gyr^{-1}}$. This mass is among the lowest inferred for host-galaxies of explosive transients from massive stars. Our observations of \sn{} are consistent with an engine-driven aspherical explosion from a rare evolutionary path of a H-rich stellar progenitor, but we cannot rule out a stellar tidal disruption event on a centrally-located intermediate mass black hole. Regardless of the physical mechanism, \sn{} establishes the existence of a new class of rare (rate $<0.4\%$ of the local core-collapse supernova rate) H-rich transients that can launch mildly relativistic outflows. 
\end{abstract}

\keywords{supernovae: individual (\sn{}) - accretion, accretion disks – stars: black holes – X-rays: general - radio: general}

%%%%%%%%%%%%%%%%%%%%%%%%%%%%%%%%%%%%%%%%%%%
\section{Introduction}
\label{Sec:intro}
Fast Blue Optical Transients (FBOTs), or alternatively Fast Evolving Luminous Transients (FELTs), are a class of transients defined by an extremely rapid rise to maximum light (typically $<12$ days), luminous optical emission ($\gtrsim 10^{43}$ erg s$^{-1}$) and blue colors. Due to their fast rise-times, they are difficult to detect and have only been identified as a class since the recent advent of high-cadence optical surveys. Only a few tens of systems have been found at optical wavelengths \citep[e.g.,][]{Matheson00, Poznanski10, Ofek10, Drout13, Drout14, Shivvers16, Tanaka16, Arcavi16, Whitesides17, Rest18, Pursiainen18,Tampo20}. Not all FBOT rise-times and luminosities can be reconciled with standard SN models \citep[e.g.,][]{Drout14}, and the diverse properties of the class have led to a range of proposed models. These include explosions of stripped massive stars \citep[e.g.,][]{Drout13,Moriya17}, shock breakout emission from an extended low-mass stellar envelope or dense circumstellar medium (CSM, e.g., \citealt{Ofek10,Drout14}), cooling envelope emission from extended stripped progenitor stars \citep[e.g.,][]{Tanaka16}, helium shell detonations on white dwarfs \citep{Shen10,Perets10}, or scenarios invoking a central engine such as a magnetar or black hole \citep[e.g.,][]{Cenko2012_PTF10iya,Hotokezaka17}. However, prior to this work, only two FBOTs (\cow{} and ZTF18abvkwla) had been detected at radio and/or X-ray wavelengths. The variable X-ray emission \citep{RiveraSandoval18}, transient hard X-ray component, steep X-ray decay and multi-wavelength evolution \citep{Margutti19} of \cow{} directly indicate a driving central engine \citep[e.g.,][]{Prentice18,Perley19,Kuin19,Margutti19,Ho19}.  
Another direct manifestation of a central engine is the presence of relativistic ejecta -- this was recently inferred for ZTF18abvkwla \citep{Ho20}.

CRTS CSS161010 J045834-081803 (hereafter referred to as \sn{}) was discovered by the Catalina Real-time Transient Survey \citep{Drake09} on 2016 October 10. The transient was also detected by the All-Sky Automated Survey for Supernovae (ASAS-SN, \citealt{Shappee14}) and showed a fast $\sim4$ day rise to maximum light at $V$-band (Dong et al., in prep.). Follow-up optical spectroscopic observations one week later showed a blue and featureless continuum \citep{Reynolds16}. These characteristics identify \sn{} as an FBOT (see \citealt{Drout14}). Further spectroscopic observations by Dong et al.\ (in prep.), showed broad spectral features (including hydrogen) and placed \sn{} at a distance of 150 Mpc ($z=0.034\pm0.001$). Optical spectroscopy of the transient host galaxy that we present here leads to $z=0.0336 \pm 0.0011$, consistent with the estimate above.

In this paper we present radio and X-ray observations of \sn{} and optical spectroscopic observations of its host galaxy. This paper is organized as follows. In  \S\ref{sec:observations} we present the observations of \sn{} and its host galaxy and in \S\ref{Sec:Radio_Xray_inferences} we infer the blast-wave properties based on the radio and X-ray observations. In \S\ref{Sec:host} and \S\ref{sec:nature} we respectively model the host properties and discuss models for \sn{}. Conclusions are drawn in \S\ref{Sec:Conc}. The optical observations and spectral evolution will be presented in Dong et al.\ (in prep). Time is reported relative to the estimated explosion date MJD 57667  (2016 October 6, Dong et al.\ in prep.).  $1\sigma$ uncertainties are reported unless stated otherwise (where $\sigma^2$ is the variance of the underlying statistical distribution). 

%%%%%%%%%%%%%%%%%%%%%%%%%%%%%%%%%%%%%%%%%%%%%%%%%%%%%%%%%%%%%%%%%%%%%%%%%
\section{Observations}\label{sec:observations}
%-------------------------------------------VLA
\subsection{VLA observations of CSS161010}
\label{SubSec:VLA}

\begin{figure} 
	\centering
	\includegraphics[width=0.5\textwidth]{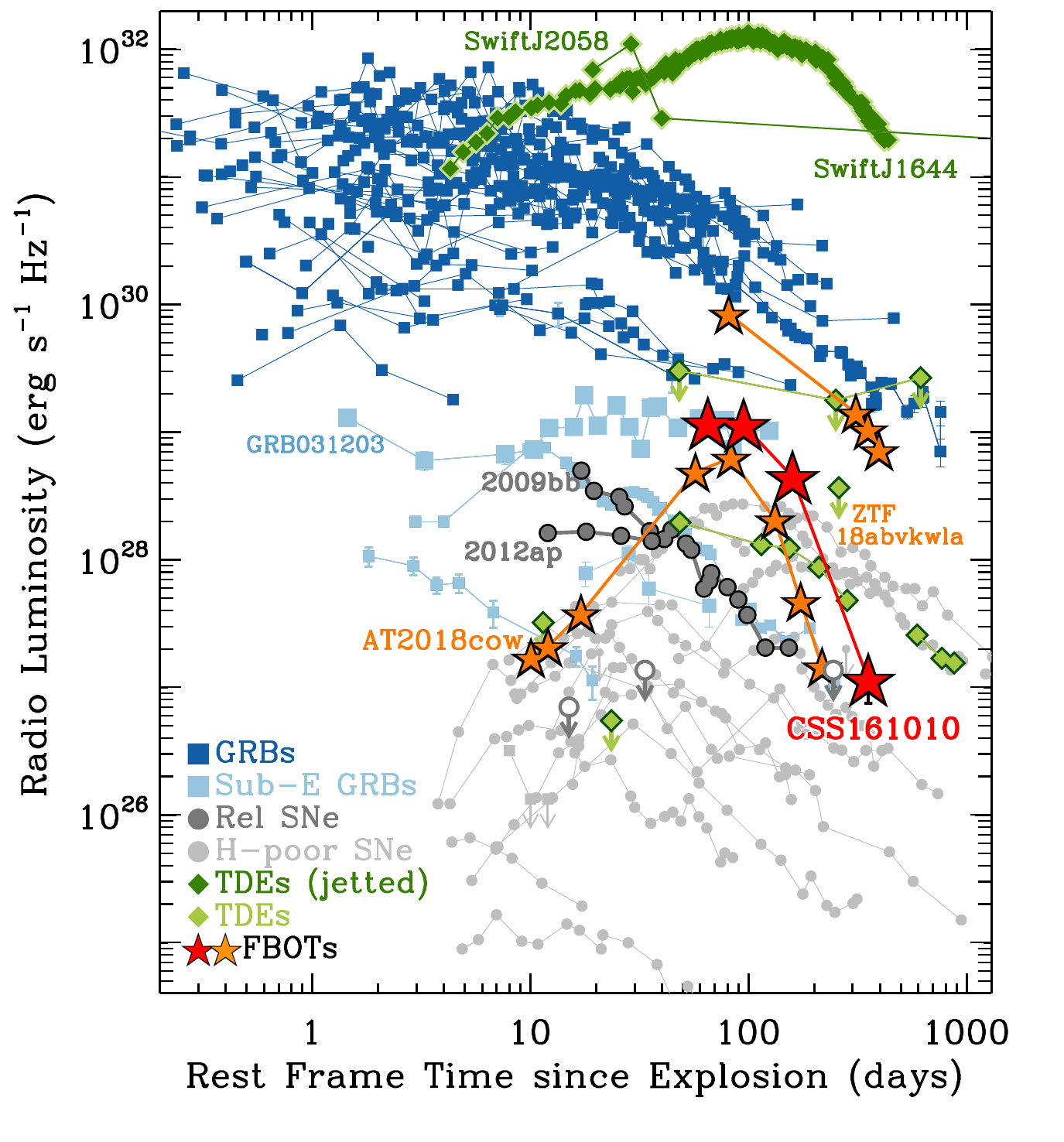}
    \caption{The 8-10 GHz light curve of \sn{} (red stars) in the context of those of other classes of explosive transients including GRBs (blue squares), sub-energetic GRBs (light-blue squares), relativistic SNe (dark grey circles), normal H-stripped core-collapse SNe (light-grey circles), TDEs (light-green diamonds) and TDEs with relativistic jets (dark-green diamonds). Empty grey circles mark the non-detection of the very rapidly declining SN-Ic 2005ek and the rapidly rising iPTF16asu, which later showed a Ic-BL spectrum \citep{Drout13,Whitesides17}. \sn{} had a radio luminosity similar to that of the sub-energetic GRB\,031203 and higher than that of relativistic SNe, normal SNe and some sub-energetic GRBs. \sn{} declined significantly more rapidly than any of these source classes, including the GRBs. The other two FBOTs with detected radio emission are also shown, with orange stars.  References: \citet{Berger12, Cenko12, Chomiuk12b, Chandra12, Zauderer13, Drout13, Chornock14,Ho19,Margutti13b, Nicholl16, Alexander16, Brown17, Margutti17b, Margutti19, Whitesides17, Mattila18, Eftekhari18, Ho20}, Coppejans et al., in prep. 
    }
    \label{Fig:RadioComparison}
\end{figure}

We observed \sn{} with the NSF's Karl G. Jansky Very Large Array (VLA) through project VLA/16B-425 and VLA/18A-123 (PI: Coppejans) over five epochs from December 2016 to March 2018, $\delta t=69-530$ days after explosion (Table \ref{Tab:radio} and Figure \ref{Fig:RadioComparison}). To monitor the spectral evolution of the source, we observed at mean frequencies of 1.497 (L-band), 3 (S-band), 6.048 (C-band), 10.0 (X-band) and 22.135 GHz (K-band). The bandwidth was divided into 64 (K-band), 32 (X-band) 16 (C-band and S-band) and 8 (L-band) spectral windows, each subdivided into 64 2-MHz channels. The observations were taken in standard phase referencing mode, with 3C147 as a bandpass and flux-density calibrator and QSO J0501--0159 and QSO J0423--0120 as complex gain calibrators.

We calibrated the data using the VLA pipeline in the Common Astronomy Software Applications package \citep[CASA,][]{McMullin07} v4.7.2., with additional flagging. For imaging we used Briggs weighting with a robust parameter of 1, and only performed phase-only self-calibration where necessary. We measured the flux density in the image-plane using PyBDSM (Python Blob Detection and Source Measurement \citealt{Mohan15}) with an elliptical Gaussian fixed to the dimensions of the CLEAN beam. To more densely sample the cm-band spectral energy distribution, we subdivided the available bandwidth into 128 MHz sections where possible and imaged each individually. We verified the pipeline reduction by undertaking manual flagging, calibration, imaging, and self-calibration of the first three epochs of VLA observations in CASA. The derived flux densities were consistent with the values measured from the VLA pipeline calibration. We report the flux densities from the VLA pipeline-calibrated data together with a more detailed description of each observation in Table \ref{Tab:radio}. The position that we derive for \sn{} from these radio observations is RA=04:58:34.396$\pm0.004$, dec=--08:18:03.95$\pm0.03$.

%-------------------------------------------GMRT
\subsection{GMRT observations of CSS161010}
\label{SubSec:GMRT}

We observed \sn{} for 10 hours with the Giant Metrewave Radio Telescope (GMRT) under the project code DDTB287 (PI: Coppejans). These observations were carried out on 2017 September 14.93, 21.96, 19.88 UT ($\delta t=344-351$ days after explosion) at frequencies 1390, 610 and 325 MHz, respectively (Table \ref{Tab:radio}). 
The 33 MHz observing bandwidth was split into 256 channels at all three frequencies. We used the Astronomical Image Processing Software (AIPS) to reduce and analyze the data. Specifically, for flagging and calibration we used the FLAGging and CALibration (FLAGCAL) software pipeline developed for GMRT data \citep{Prasad12}. Additional manual flagging and calibration was also performed. We performed multi-facet imaging to 
to deal with the field which is significantly curved over the GMRT field-of-view. The number of facets was calculated using the \texttt{SETFC} task. Continuum images were made using the \texttt{IMAGR} task. For each observation we performed a few rounds of phase-only self calibration and one round of amplitude and phase self-calibration. The errors on the flux density were calculated by adding those given by the task \texttt{JMFIT} and a 15\% systematic error in quadrature.

The source positions in our GMRT and VLA images are consistent.
To compare the flux density scaling of the VLA and GMRT data, we took an observation at $\sim1.49$ GHz with each telescope (these observations were separated by two weeks and the central frequencies differed by 0.107 GHz) and the flux densities were consistent. Additionally, we confirmed that the flux density of a known point source in our GMRT 1.4 GHz image was consistent with that quoted in the National Radio Astronomy Observatory (NRAO) VLA Sky Survey (NVSS; \citealt{Condon98}) source catalogue.

%-------------------------------------------Chandra
\subsection{Chandra observations of CSS161010}
\label{SubSec:CXO}
\begin{figure} 
	\centering
	\includegraphics[width=0.5\textwidth]{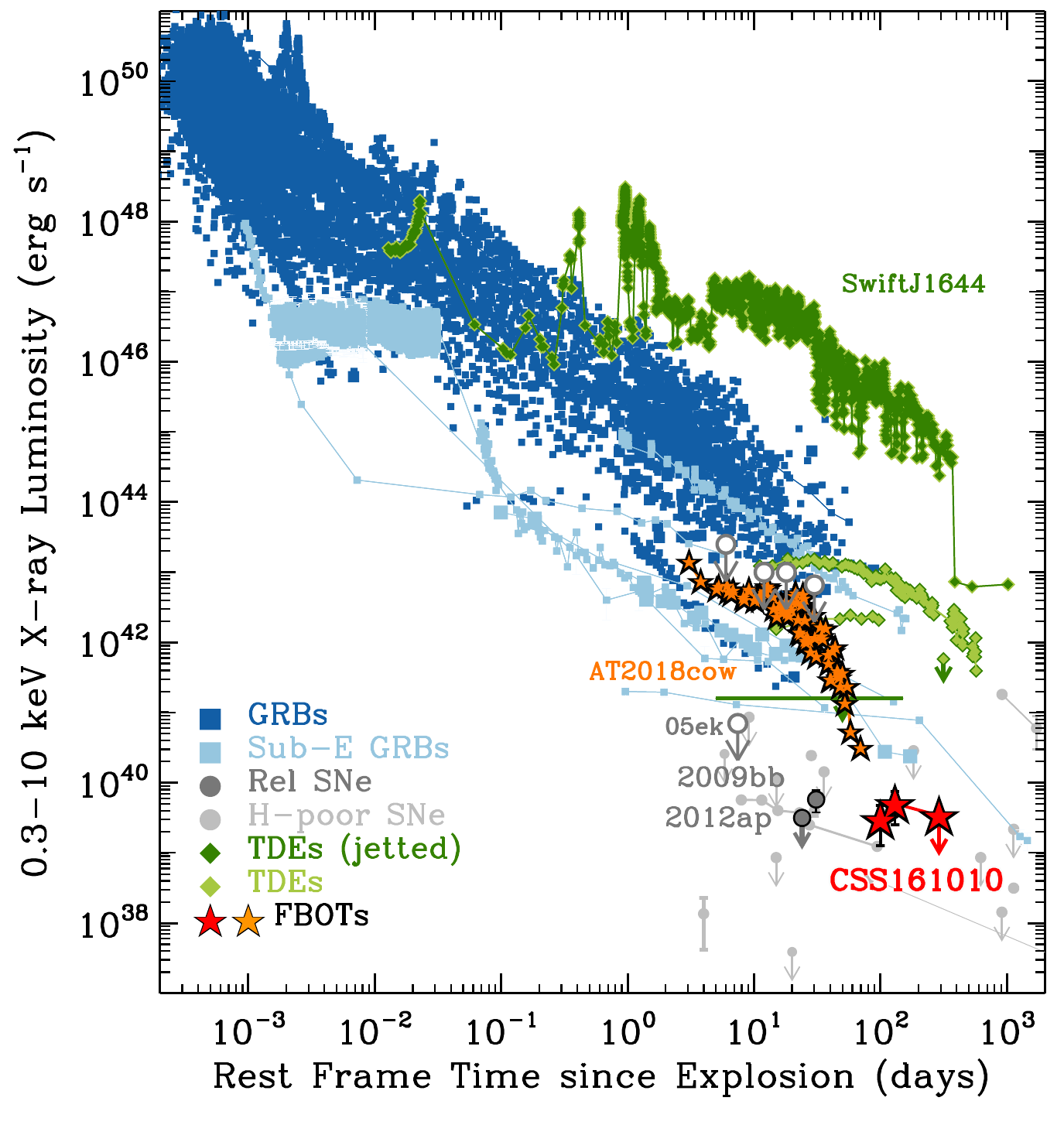}
    \caption{The 0.3-10 keV X-ray luminosity of \sn{} in the context of those of other classes of transients following the same color scheme as in Fig. \ref{Fig:RadioComparison}.  \cow\ is the only other FBOT with detected X-ray emission. Empty circles mark the upper limits on the X-ray luminosities of the very rapidly declining type-Ic SN\,2005ek, the rapidly rising iPTF16asu, which later showed a Ic-BL spectrum, and the fast-rising, luminous transient ``Dougie''. References: \cite{Margutti13b,Margutti13,Drout13,Vinko15,Whitesides17,Margutti19}.}
    \label{Fig:XrayComparison}
\end{figure}

We initiated deep X-ray observations of \sn{} with the Chandra X-ray Observatory (CXO) on January 13, 2017 under a DDT program (PI Margutti; Program 17508566; IDs 19984, 19985, 19986). Our CXO observations covered the time range $\delta t\sim99-291$ days after explosion (Fig. \ref{Fig:XrayComparison}). The ACIS-S data were reduced with the {\tt CIAO} software package (v4.9) and relative calibration files, applying standard ACIS data filtering. A weak X-ray source is detected at the location of the optical transient in our first two epochs of observation at $t\sim99$ days and $\sim 130$ days, while no evidence for X-ray emission is detected at $\sim291$ days. 

In our first observation (ID 19984, exposure time of 29.7 ks) we detect three photons in a 1\arcsec\ region around the transient, corresponding to a 3.9$\,\sigma$ (Gaussian equivalent) confidence-limit detection in the 0.5-8 keV energy range, at a  count-rate of $(1.01\pm 0.58)\times 10^{-4}\,\rm{c\,s^{-1}}$ (the uncertainty here reflects the variance of the underlying Poissonian process). For an assumed power-law spectrum with photon index $\Gamma=2$ and no intrinsic absorption, the corresponding unabsorbed 0.3-10 keV flux is $F_x=(1.33\pm0.76)\times 10^{-15}\,\rm{erg\,s^{-1}cm^{-2}}$ and the luminosity is $L_x=(3.4\pm 1.9)\times 10^{39}\,\rm{erg\,s^{-1}}$. The Galactic neutral hydrogen column density in the direction of the transient is $NH_{\rm MW}=4.7\times 10^{20}\,\rm{cm^{-2}}$ \citep{Kalberla05}. 

The X-ray source is still detected at the location of \sn{} at the time of our second CXO observation on February 13, 2017 (ID 19985, exposure time of 27.1 ks), with a count-rate of $(1.48\pm 0.74)\times 10^{-4}\,\rm{c\,s^{-1}}$ and significance of 4.7$\sigma$ (0.5-8 keV). The corresponding unabsorbed flux is $F_x=(1.94\pm0.97)\times 10^{-15}\,\rm{erg\,s^{-1}cm^{-2}}$ (0.3-10 keV), and the luminosity is $L_x=(5.0\pm2.5)\times 10^{39}\,\rm{erg\,s^{-1}}$. 
 
The X-ray emission had faded by the time of our third observation on July 23, 2017 (ID 19986, exposure time of 29.4 ks) and we place a 3$\sigma$ count-rate upper limit $<1.02\times 10^{-4}\,\rm{c\,s^{-1}}$ (0.5-8 keV), which corresponds to $F_x<1.31\times 10^{-15}\,\rm{erg\,s^{-1}cm^{-2}}$ and $L_x<3.4\times 10^{39}\,\rm{erg\,s^{-1}}$ (0.3-10 keV).

%--------------------------------------------
\subsection{Constraints on the Prompt $\gamma$-ray Emission}
\label{sec:gammarays}

We searched for associated prompt $\gamma$-ray emission from \sn{} around the time of explosion with the Inter-Planetary Network (IPN; Mars Odyssey, Konus/Wind, INTEGRAL SPI-ACS, Swift-BAT, and Fermi-GBM). Based on the optical photometry of the rise (Dong et al.\ in prep.), we used a conservative explosion date of JD=$2457669.7\pm2$ for this search. We estimate an upper limit (90\% conf.) on the 20 - 1500 keV fluence of $\sim8\times10^{-7}$ erg~cm$^{-2}$ for a burst lasting less than 2.944 s and having a typical Konus Wind short GRB spectrum (an exponentially cut off power law with $\alpha=-0.5$ and $E_p=500$ keV). For a typical long GRB spectrum (the Band function with $\alpha=-1$, $\beta=-2.5$, and $E_p=300$ keV), the corresponding limiting peak flux is $\sim$$2$$\times10^{-7}$ erg cm$^{-2}$ s$^{-1}$ (20-1500 keV, 2.944 s scale). The peak flux corresponds to a peak luminosity  $L_{pk}<5\times10^{47}\,\rm{erg\,s^{-1}}$. For comparison, the weakest long GRBs detected have $L_{pk}\approx10^{47}\,\rm{erg\,s^{-1}}$ (e.g. \citealt{Nava12}).

%-------------------------------------------Host Galaxy
\subsection{Host Galaxy Observations}
\label{SubSec:host}

\begin{figure*} 
	\centering
	\includegraphics[width=1\textwidth]{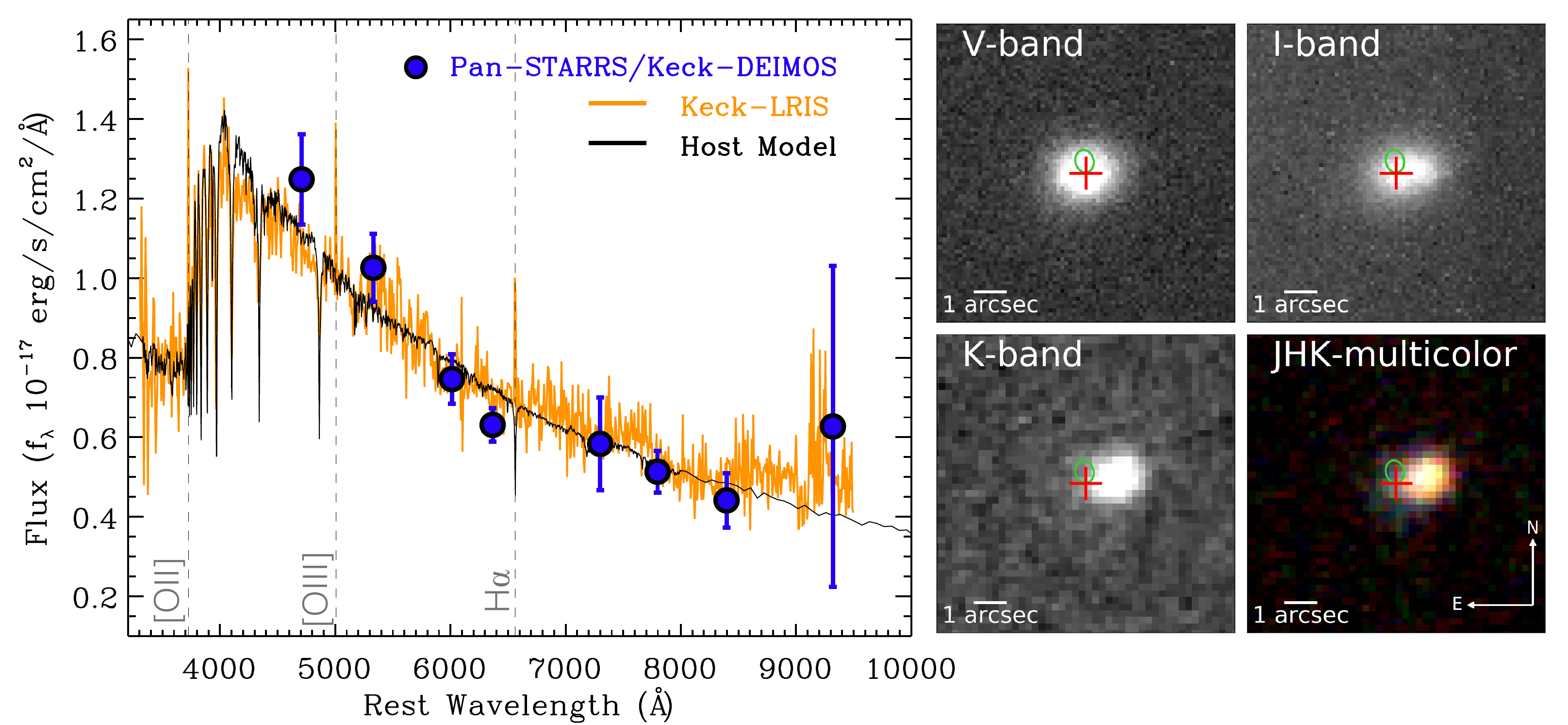}
    \caption{Spectrum, photometry and images of the dwarf galaxy host of CSS161010 at $z=0.0336\pm0.0011$. \emph{Left Panel:} The orange line shows the spectrum of \sn's host as observed by Keck/LRIS, while the blue points show the Pan-Starrs/Keck-DEIMOS measurements.  The observations have been corrected for a Galactic extinction of $E(B-V)=0.084\,\rm{mag}$. The Keck/LRIS spectrum has been re-scaled to the Pan-STARRS and Keck/DEIMOS photometry (blue filled circles) as part of the fitting procedure. The black line shows the best-fit FAST model, which has a total stellar mass of $\sim10^7\,M_{\odot}$ and current star-formation rate $\sim0.004\,\rm{M_{\odot}\,yr^{-1}}$. \emph{Right Panels:} Optical (\textit{V}- and \textit{I}-band from Keck-DEIMOS) and NIR (\textit{JHK}-bands from MMT+MMIRS) images of the surroundings of \sn{}. The red cross marks the position of the centroid of the dwarf host galaxy visible in \textit{V}-band and the green ellipse marks the 5$\sigma$ contour of the radio transient at 6 GHz, which is consistent with the optical position of the transient. The apparent shift of the centroid of the emission in the redder bands is due to contamination by a red source (possibly a red dwarf star) almost coincident with the position of the host galaxy of \sn{}. The radio emission is not associated with the contaminating red source.}
    \label{Fig:host}
\end{figure*}

\sn{} has a faint host galaxy that is visible in deep optical images of the field. The location of \sn{} is consistent with the inferred center of the host galaxy (RA=04:58:34.398 and dec=--08:18:04.337, with a separation of 0\farcs39). We acquired a spectrum of this anonymous host galaxy on 2018 October 10 ($\delta t=$790 days since explosion) well after the optical transient had completely faded away. We used the Keck Low Resolution Imaging Spectrometer (LRIS) equipped with the 1\farcs0 slit, the 400/3400 grism for the blue side (6.5~\AA\ resolution) and the 400/8500 grating for the red side (6.9~\AA\ resolution), covering the wavelength range between 3400 and 10200~\AA, for a total integration time of 3300~s. The 2-D image was corrected for overscan, bias and flatfields, and the spectrum was then extracted using standard procedures within \texttt{IRAF}\footnote{http://iraf.noao.edu/}. The spectrum was wavelength and flux calibrated using comparison lamps and a standard star observed during the same night and with the same setup. A Galactic extinction $E(B-V)=0.084$ mag in the direction of the transient was applied \citep{Schlafly11}. 

On 2019 February 25, we imaged the field of the host galaxy of \sn{} in the \textit{VRI} optical bands with Keck+DEIMOS, using an integration time of 720~s for each filter. We used SExtractor \citep{Bertin1996} to extract the isophotal magnitudes of the host galaxy of \sn{}. We calibrated this photometry using the fluxes of the field stars retrieved from the Pan-STARRS1\footnote{https://panstarrs.stsci.edu} catalogue \citep{Chambers16}. We converted the \textit{gri} magnitudes of the Pan-STARRS1 field stars to Johnson/Cousins \textit{VRI} magnitudes following \citet{Chonis2008}. The final Vega magnitudes of the host of \sn{} are $V = 21.68\pm0.09$ mag, $R = 21.44\pm0.07$ mag, $I = 20.91\pm0.08$ mag. We then used the same technique to extract $gri$ magnitudes of the host from the Pan-STARRS1 data archive images of $g=21.9\pm0.1$ mag, $r=21.1\pm0.1$ mag and $i=20.6\pm0.1$ mag.

We obtained near-infrared (NIR) imaging of the field of \sn{} with MMT and the Magellan Infrared Spectrograph \citep[MMIRS][]{2012PASP..124.1318M} in imaging mode on 2018 November 15. We acquired \textit{JHK} images with 60~s exposures for a total integration time of 900~s for \textit{J}, and 1200~s for \textit{H} and \textit{K}. We processed the images using the MMIRS data reduction pipeline \citep{2015PASP..127..406C}.
A separate NIR source is clearly detected at RA=04:58:34.337 dec=--08:18:04.19, 0\farcs91 from the radio and optical location of \sn{} (Fig.\ \ref{Fig:host}). This source dominates the NIR emission at the location of \sn. The inferred Vega measured magnitudes of this contaminating source calibrated against the Two Micron All Sky Survey (2MASS) catalog\footnote{\url{http://www.ipac.caltech.edu/2mass/}} \citep{Skrutskie06} are $J = 19.24\pm0.30$ mag, $H = 18.09\pm0.08$ mag, $K = 17.76\pm0.11$ mag. We note that this source is also detected in the WISE (Wide-field Infrared Survey Explorer) W1 and W2 bands. To measure WISE W1 (3.4$\mu$m) and W2 (4.6$\mu$m) fluxes, we performed
PSF photometry on the \cite{meisner2018} unWISE coadds. These stacks have a $\sim 4 \times$ greater depth than AllWISE,
allowing for higher S/N flux measurements. We infer Vega magnitudes of $W1$
= 16.94 $\pm$ 0.07 and $W2$ = 16.74 $\pm$ 0.17. The uncertainties were estimated via PSF fitting of Monte Carlo image realizations with an
appropriate per-pixel noise model.
According to \citet{Jarrett17}, $W1-W2=0.2\pm0.2$ mag rules out active galactic nuclei, T-dwarfs and ultra luminous infrared galaxies. This contaminating source is therefore most likely a foreground star.

%%%%%%%%%%%%%%%%%%%%%%%%%%%%%%%%%%%%%%%%%%%
\section{Inferences from the Radio and X-ray observations}
\label{Sec:Radio_Xray_inferences}

\subsection{Radio Spectral Evolution and Modelling}\label{sec:radio_spectrum}

\begin{figure*}
	\centering
	\includegraphics[width=0.95\textwidth]{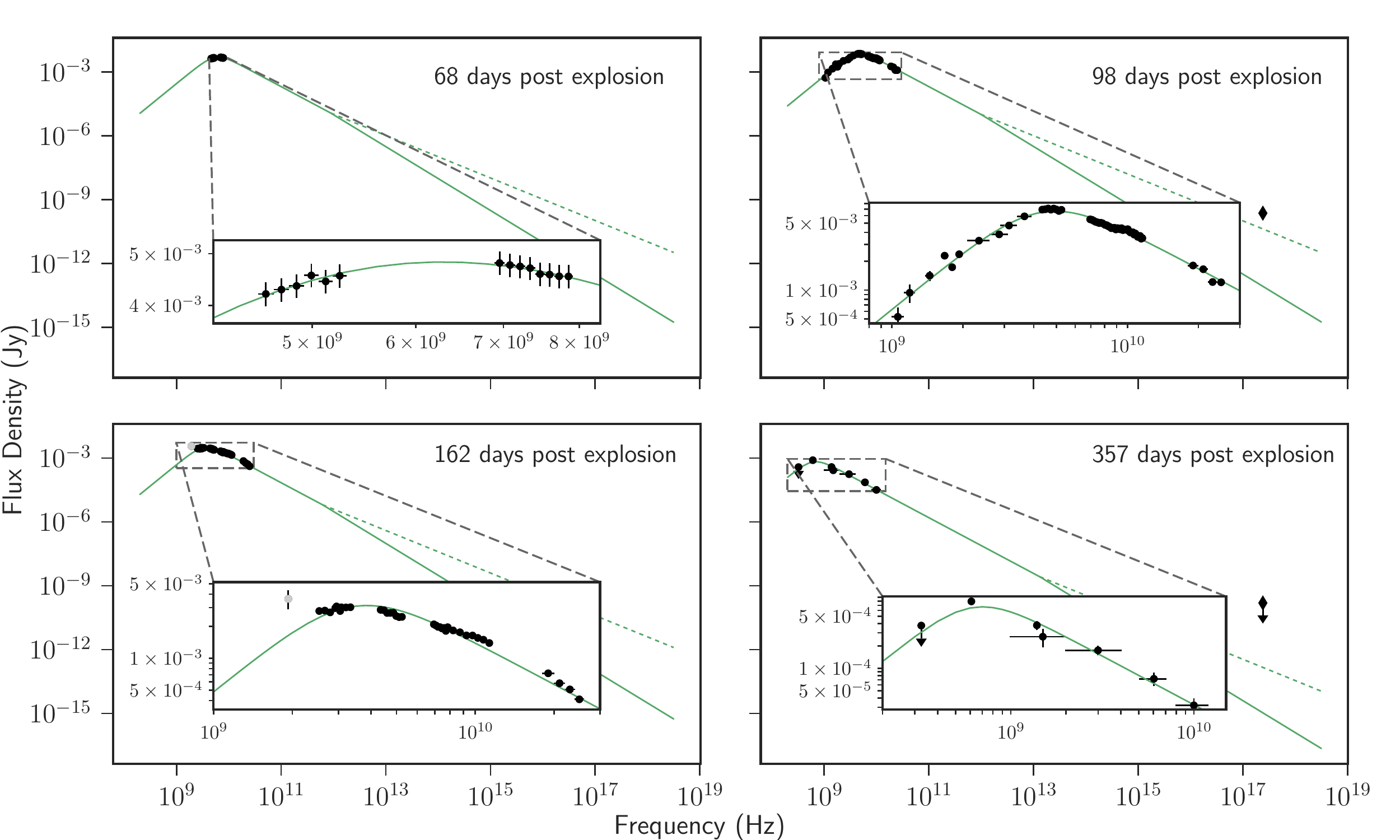}
    \caption{Broad-band radio-to-Xray spectral energy distribution of \sn{} (black points) along with fits with a synchrotron self-absorbed (SSA) model (green line). The smoothing parameter ($-0.9$), and optically thick ($2.00\pm0.08$) and thin ($-1.31\pm0.03$) spectral indices for the fits were derived from our most constraining epoch, at 99 days post explosion. The fitted values are given in Table~\ref{Tab:properties}. The measurements below 2 GHz at 162 days post explosion were strongly affected by radio frequency interference and we flagged out much of this band. Subsequently, we treat the lowest frequency point (shown in light gray) with caution. The X-ray emission does not fall on the same SSA spectrum, as the spectral index steepens at frequencies above the cooling break. The dotted (green) line shows the extrapolation of the SSA spectrum without taking the cooling break into account. Note that the X-ray observation in the bottom right panel was taken at 425 days post explosion.
    }
    \label{Fig:SED_fits_SSA}
\end{figure*}

The observed radio spectral evolution is consistent with a synchrotron self-absorbed (SSA) spectrum where the self-absorption frequency $\nu_{\rm sa}$ evolves to lower frequencies as the ejecta expands and becomes optically thin (Fig.\ \ref{Fig:SED_fits_SSA}). The optically thick and thin spectral indices derived from our best-sampled epoch (99 days post explosion) are $\alpha=2.00\pm0.08$ and $\alpha=-1.31\pm0.03$, respectively (where $F_{\nu}\propto\nu^{\alpha}$). The optically thin flux density scales as $F_{\nu}\propto \nu^{-(p-1)/2}$, where $p$ is the index of the distribution of relativistic electrons responsible for the synchrotron emission $N_e\propto (\gamma_e)^{-p}$ and $\gamma_e$ is the Lorentz factor of the electrons (we find $p=3.6^{+0.4}_{-0.1}$). Table \ref{Tab:properties} and Figure \ref{Fig:SED_fits_SSA} show the peak frequency $\nu_p$ (which is equivalent to the self-absorption frequency $\nu_{\rm sa}$), the peak flux density ($F_p$) and the parameters derived for the SSA spectrum by fitting each epoch with a broken power-law.
We find $\nu_p\propto t^{-1.26\pm0.07}$ and $F_p\propto t^{-1.79\pm0.09}$ a steep decay in the radio luminosity of $L_{8\,\rm{GHz}}\propto t^{-5.1\pm0.3}$ at $>99$ days post explosion). The evolution of the SSA peak is consistent with an expanding blast-wave, but is different from the evolution of a SSA-dominated, non-strongly decelerating, non-relativistic SN in a wind-like medium where $\nu_p\propto t^{-1}$ and $F_p\sim$ constant \citep[][]{Chevalier98,Soderberg05,Soderberg06d}. The inferred $F_p(t)$ is also steeper than seen in relativistic SNe (see \S\ref{sec:dirty_fireball}). We compare these properties to the two other radio-detected FBOTs in \S\ref{sec:FBOTs}.

The physical properties of an expanding blastwave can be calculated from an SSA spectrum if $F_p$, $\nu_p$, the source distance, and the fractions of energy in the relativistic electrons ($\epsilon_e$) and magnetic fields ($\epsilon_B$) in the internal shock are known \citep{Scott1977,Slysh90,Readhead94,Chevalier98,Chevalier06}. We follow the SSA modelling framework for SNe \citep{Chevalier98,Chevalier06} to obtain robust estimates of the blastwave radius $R$ and velocity, environment density $n$, internal energy $U_{\rm int}$ and magnetic field $B$.
We employ the subscript `eq' to identify quantities derived under the assumption of equipartition (i.e., $\epsilon_e=\epsilon_B=1/3$). We emphasize that our estimates of $B$ and $R$ (and subsequently the shock velocity) are only \textit{weakly} dependent on the microphysical parameters. The normalizations of $U_{\rm int}$ and $n$ do depend on the shock microphysics, but the inferred variation of these parameters with time does not. We do not assume any time-dependent evolution for the blastwave, but rather fit each epoch individually to derive the blastwave properties given in Table \ref{Tab:properties}. The relations quoted below were obtained by fitting a power-law to these properties over the epochs at 69, 99 and 357 days post explosion. Our major conclusions are not affected if we include our least constrained epoch (162 days post explosion) in the fits.

%%%%%%%%%%%%%%%%%%%%%%
\subsection{A Mildly Relativistic, Decelerating Blast-wave in a Dense Environment}\label{sec:mildly_rel}

\begin{table*}
\centering
    \caption{Radio spectral properties and derived blast-wave properties.}
    \label{Tab:properties}
    \begin{tabular}{ccccccccc}
    \hline
Time$^{\rm{a}}$ & $\nu_{p}$ $^{\rm{b}}$ & $F_{p}$ $^{\rm{b}}$ & $R_{\rm eq}$ & $B_{\rm eq}$ & $(\Gamma\beta c)_{\rm eq}$ $^{\rm{c}}$ & $U_{\rm eq}$ & $n_{\rm eq}$ & $\dot{M}_{\rm eq}$ $^{\rm{d}}$\\
(days) & (GHz) & (mJy) & ($10^{16}$ cm) & (G) & (c) & ($10^{49}$ erg) & (cm$^{-3}$) & ($10^{-5}\,M_{\odot}y^{-1}$)\\ 
\hline
69 & $5.6\pm0.2$ & $8.8\pm0.2$ & $9.5\pm0.4$ & $0.38\pm0.02$ & $0.53\pm0.02$ & $2.9\pm0.1$ & $47\pm5$ & $1.4\pm0.1$\\ 
99 & $4.4\pm0.1$ & $12.2\pm0.3$ & $14.1\pm0.5$ & $0.29\pm0.01$ & $0.55\pm0.02$ & $5.6\pm0.2$ & $25\pm2$ & $1.7\pm0.1$\\ 
357 & $0.63\pm0.07$ & $1.2\pm0.1$ & $33\pm4$ & $0.052\pm0.006$ & $0.36\pm0.04$ & $2.4\pm0.4$ & $1.9\pm0.6$ & $0.7\pm0.2$\\
\hline
\end{tabular}
\tablecomments{\footnotesize{$^{\rm{a}}$ As the observations at 162 days were strongly affected by radio frequency interference at low frequencies and we had to flag most of the data (Figure~\ref{Fig:SED_fits_SSA}), the optically thick emission was not constrained and we do not include the results for this epoch here or in our modelling. For reference, the derived parameters at 162 days are $F_p=3.4\pm0.1$, $\nu_p=5.8\pm0.1$, $R_{\rm eq}=12.7\pm0.5$, $B_{\rm eq}=0.241\pm0.009$, $(\Gamma\beta c)_{\rm eq}=0.30\pm0.01c$, $U_{\rm int,eq}=2.9\pm0.1$, $n_{\rm eq}=59\pm6$ and $\dot{M}_{\rm eq}=3.2\pm0.2$. $^{\rm{b}}$ Frequency (column 2) and flux density (column 3) at the intersection of the optically thin and thick synchrotron power-laws, from which we calculate the blast-wave parameters following \cite{Chevalier98}. $^{\rm{c}}$ Average apparent velocity $(\Gamma\beta c)_{\rm eq}c=R_{\rm eq}c/t$. $^{\rm{d}}$ For wind velocity $v_w=$1000 $\rm{km\,s^{-1}}$.}}
\end{table*}

Over the 308 days spanned by our observations the forward shock radius in \sn{} expanded according to $R=3\times10^{15}(f\epsilon_e/\epsilon_B)^{-1/19}(t_{\rm obs}/\rm{days})^{0.81\pm0.08}$ cm, where $R$ is calculated from Equation 21 in \cite{Chevalier06}, $f$ is the fraction of the spherical volume producing radio emission,
and $t_{\rm obs}$ is the time since explosion.  In the absence of strong relativistic beaming (which applies to Lorentz factors $\Gamma\gg 1$), the radio emission effectively provides a measure of the blastwave lateral expansion (instead of the radius along our line of sight) or $(\Gamma\beta)c=R/t_{\rm obs}$, from which we derive an apparent transverse velocity up to 99 days (our 
best-constrained epoch) of $(\Gamma\beta c)_{\rm eq}=0.55\pm0.02c$. The blastwave was decelerating during our observations, as at 357 days post explosion we measured $(\Gamma\beta c)_{\rm eq}=0.36\pm0.04c$. Because of the equipartition assumption and the deceleration of the blastwave, we conclude an initial $\Gamma\beta c>0.6c$. This result implies a decelerating, mildly relativistic blastwave, with similarities to the radio-loud FBOT event ZTF\,18abvkwla (\S\ref{sec:FBOTs}, \citealt{Ho20}). We thus conclude that \sn{} is an FBOT with a mildly relativistic, decelerating outflow, and is the first relativistic transient with hydrogen in its ejecta (optical spectroscopic observations presented in Dong et al., in prep.). 

Following the standard \citet{Chevalier06} framework for synchrotron emission from SNe,  we further derive  an environment density profile $n=12\epsilon_B^{-1}(\epsilon_ef/\epsilon_B)^{-8/19}(r_{\rm}/10^{17}\,\rm{cm})^{-2.3\pm0.3}\,\rm{cm}^{-3}$ at $r_{\rm eq}\geq9.5\times10^{16}$ cm.  For fiducial microphysics values ($f\approx0.5, \, \epsilon_e=0.1, \, \epsilon_B=0.01$) this result implies $n\approx700\,\rm{cm^{-3}}$ at $r\approx10^{17}$ cm, corresponding to an effective mass-loss rate of $\dot{M}\approx2\times10^{-4}\,\Msol\,y^{-1}$ for a wind velocity of 1000 $\rm{km\,s^{-1}}$. The inferred environment density is fairly high for massive stars \citep[see][]{Smith14} and comparable to the densities inferred for \cow{} ($\dot{M}\sim10^{-4}-10^{-3}\,\Msol\,y^{-1}$ \citealt{Margutti19}). However, \cow{} has a non-relativistic blastwave with $v\sim0.1c$ and limited (if any) deceleration over the first 150 days \citep{Ho19,Margutti19, Bietenholz20}. 

\emph{If} \sn{} originated from a massive stellar explosion (see \S\ref{sec:nature} for discussion) and the radio emission was powered by the interaction of the entire outer stellar envelope with density profile $\rho_{\rm{SN}}\propto r^{-q}$ with the medium of density $\rho_{\rm{CSM}}\propto r^{-s}$, we would expect the transient to be still in the ``interaction'' regime during the time of our radio observations (e.g. \citealt{Chevalier82b}). During this phase the shock radius expands as $R\propto t^m$ with $m=(q-3)/(q-s)$ \citep{Chevalier82b}, which implies $q\sim7$ with $s = 2$. It is unclear if the entire outer envelope is contributing to the radio emission, or if, instead, the radio-emitting ejecta constitutes a separate ejecta component (as in long GRBs, which have a relativistic jet and a spherical non-relativistic ejecta component associated with the SN). It is thus possible that \sn{} was already in the energy-conserving limit at $t\sim100$ days. We discuss below our inferences in this limit.

In the non-relativistic energy-conserving regime the Sedov-Taylor solution applies (ST, \citealt{vonNeumann41,Sedov46,Taylor50}) and the shock position scales  as $R\propto t^{2/(5-s)}$, from which we would derive $s \sim 2.5$. In the ultra-relativistic $\Gamma\gg1$ energy-conserving limit the Blandford-McKee (BM) solution \citep{Blandford76} applies, $\Gamma\propto R^{(s-3)/2}$ and $dt_{\rm obs}\sim 2dt/\Gamma^2$, from which  $R\propto t_{\rm obs}^{1/(4-s)}$, leading to $s \sim 2.7$.\footnote{In the discussion of relativistic effects we distinguish  between observed time $t_{obs}$ and time in the frame where the blastwave is spherical $t$. Everywhere else $t$ stands for time in the observer frame.} The non-relativistic and ultra-relativistic limits, both of which are self-similar, suggest a steep density profile. However, the mildly relativistic nature of the outflow of \sn{} implies that the blastwave expansion is fundamentally not self-similar, as the speed of light contributes an additional velocity scale that characterizes the expansion of the blastwave (i.e., a velocity scale in addition to the non-relativistic, energy-conserving velocity scaling $V^2 \propto R^{s-3}$).  We therefore do not expect the shock position to behave as a simple power-law with time, but to instead show some degree of secular evolution as the blast transitions to the non-relativistic regime in which the dependence on the speed of light is lost.

For mildly relativistic shocks we expect the standard ST scaling to hold up to terms that are proportional to $V^2/c^2$; \cite{Coughlin19} showed that the coefficient of proportionality multiplying this correction, $\sigma$, is a parameter that depends on the post-shock adiabatic index of the gas (effectively equal to 4/3) and the ambient density profile (see their Table 1). In particular, following \citealt{Coughlin19} (their Equation 51), in the mildly relativistic regime the shock velocity varies with position as 
\begin{equation}
R^{3-s}\Gamma^2 V^2=V_i^2(1+\sigma V^2/c^2), 
\label{Eq:MR}
\end{equation}
where $V_i$ is the velocity that the shock would have if we ignored relativistic corrections and the shock position is normalized to the time at which the shock sweeps up a comparable amount of rest mass to the initial mass. Inverting and integrating Equation \ref{Eq:MR} and accounting for $dt_{\rm obs}=(1-\beta \cos\theta)dt$ (for a patch of the shell at an angle $\theta$ with respect to the observer line of sight), it is possible to determine $R(t_{\rm obs})$. An additional complication in the mildly relativistic regime is that the observed emitting surface is viewed at delayed times for different $\theta$; specifically, photons arriving from the poles were radiated earlier than those emitted at the equator (in order to be observed simultaneously) when the ejecta was more relativistic and the radiation was more highly beamed out of our line of sight. Taking the two limiting cases, $dt_{\rm obs} = \left(1-V/c\right)dt$ and $dt_{\rm obs} = dt$, which apply to the early and late-time evolution, respectively, we find that the environment around \sn{} was likely steeper than those created by a constant mass-loss rate ($s=2$), and falls in between the limits provided by the ultra- and non-relativistic regimes. There is some precedent for this non-steady mass-loss. Recent observations of a number of SNe show eruptions in the centuries prior to explosion \citep[e.g., ][]{Smith14,Margutti14,Margutti17,Milisavljevic15}, and \cow{} shows a similarly steep density profile \citep{Margutti19} to \sn{}. We note that within our framework, a steeper density profile implies that the magnetic field also scales more steeply than the traditional wind scaling of $B\propto R^{-1}$. 

\subsection{Inferences on the Initial Blastwave Properties}
\label{sec:dirty_fireball}

\begin{figure*} 
	\centering
	\includegraphics[width=0.8\textwidth]{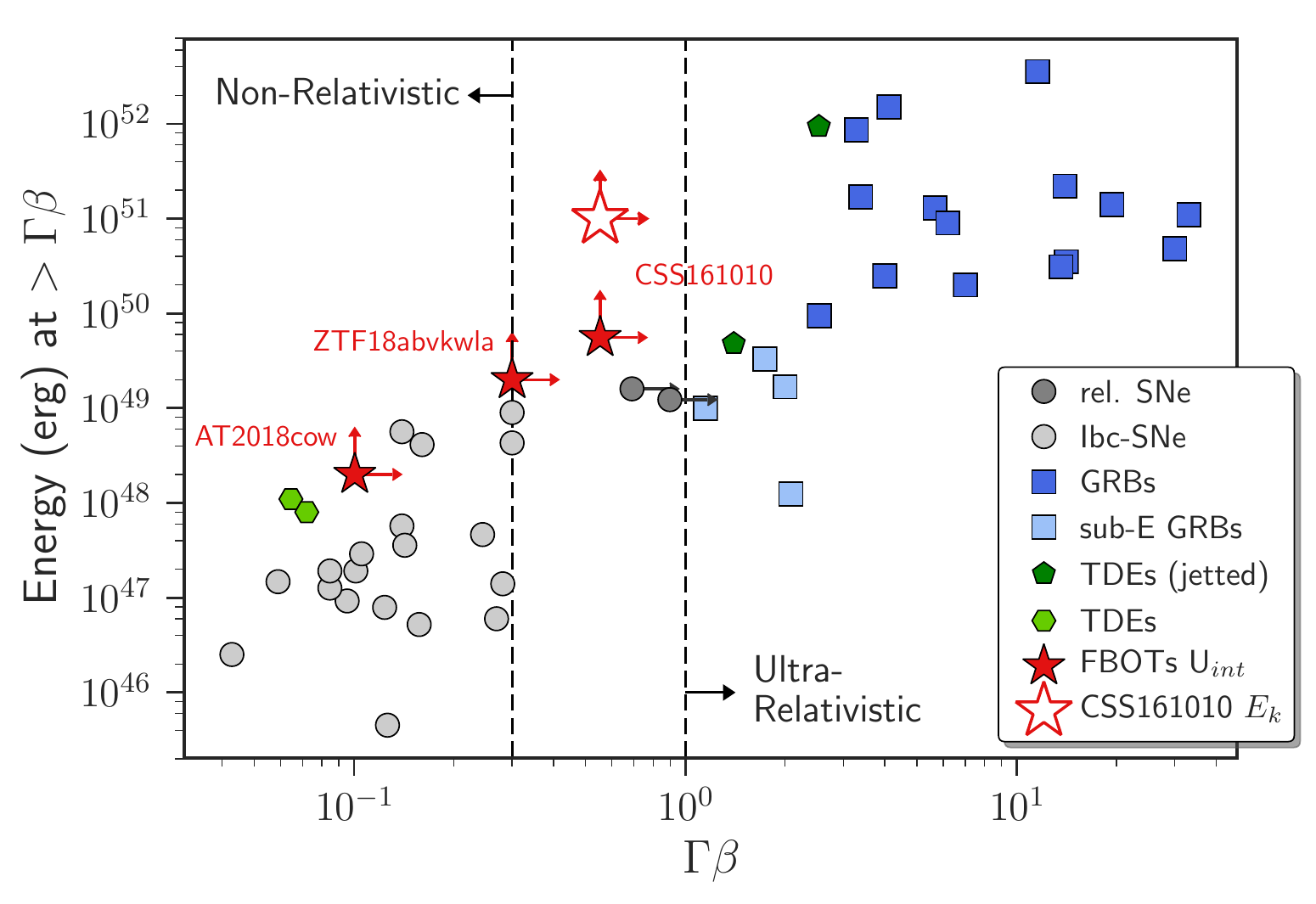}
    \caption{Kinetic energy of the fast-moving material in the outflow with velocities $>\Gamma\beta$ for \sn{} and other classes of transients, as determined from radio observations. With the exception of the FBOTs, these properties are measured at approximately 1 day post explosion. We plot the internal energy in the shock ($U_{\rm int}$) for the FBOTs at the time of the observations. For \koala{} we calculated this assuming that the 10 GHz measurement at 81 days from \citep{Ho20} is the peak of the SSA spectrum as they find a spectral index of $-0.16\pm0.05$. For \sn{} we also plot the kinetic energy at 99 days post explosion/disruption (our best constrained epoch, see Table \ref{Tab:properties}). The latter is a robust lower limit for the initial kinetic energy. \sn{} is mildly relativistic and has a velocity at least comparable to that of the relativistic SNe 2009bb \citep{Soderberg10} and 2012ap \citep{Margutti14b,Chakraborti15}. \sn{} has the fastest outflow of the FBOTs detected to date. References: \cow{} \citep{Margutti19}, \koala{} (see footnote \ref{footnote:koala}, \citealt{Ho20}), TDEs \citep{Zauderer11,Berger12,Cenko12,Alexander16,Alexander17}, GRBs and SNe \citep{Margutti13b,Margutti14} and references therein.}  
    \label{Fig:EnergyProfile}
\end{figure*}

We determined the shock internal energy $U_{int}$ at each epoch following \citet{Chevalier98}, their Equations 21 and 22. At 99 days, the equipartition conditions give a robust lower limit of $U_{\rm int}\gtrsim6\times10^{49}$ erg (Table \ref{Tab:properties}), which implies a kinetic energy of $E_{k}\gtrsim6\times10^{49}$ erg coupled to material with velocity $\Gamma\beta c\ge0.55c$. We compare the shock properties of \sn{} to those of SNe, FBOTs and TDEs in Fig.\ \ref{Fig:EnergyProfile}. The $E_k$ of the fast material in \sn{} is larger than in normal core-collapse SNe, relativistic SNe\footnote{A class of stellar explosions that show mildly relativistic outflows but no detected higher energy $\gamma$-ray counterparts (GRBs) associated with relativistic jets \citep{Soderberg10b,Margutti14,Chakraborti15,Corsi17}.},  and sub-energetic GRBs, but comparable to GRBs and relativistic TDEs. The shock powering the non-thermal emission in \sn{} is also significantly faster than in normal SNe, especially considering that it is decelerating and we are measuring it at a much later phase ($\approx99$ days post explosion) than the SNe shown in Figure \ref{Fig:EnergyProfile} at $\approx1$ day post explosion.

To estimate the initial explosion parameters, we need to extrapolate backwards by assuming a set of blast-wave dynamics. Since the early evolution of the blastwave at $t<70$ days is not constrained by our observations  we proceed with robust order-of-magnitude inferences. As the blast-wave expands and interacts with the surrounding medium its $E_k$ is converted into $U_{\rm int}$, which implies that the shock's initial $E_k$ is $E_{k,0}$$>$$U_{\rm int}$ or $E_{k,0}>10^{50-51}$ erg for fiducial values $\epsilon_e=0.1$ and $\epsilon_B=0.01$. The fact that the shock is decelerating means that the swept-up CSM mass is comparable to or exceeds the mass of the fast material in the blast-wave. We can thus estimate the fast ejecta mass and kinetic energy. During our observations the shock wave swept up $M_{\rm sw}\sim10^{-2}\,\Msol$ ($\sim10^{-3}\,\Msol$ in equipartition) as it expanded from $1\times10^{17}$~cm to $3\times10^{17}$~cm. The density profile at smaller radii is not constrained, but for profiles ranging from flat to $r^{-2.3}$ we derive a total swept up mass of $M_{\rm sw}\sim0.01-0.1\,\Msol$ ($M_{\rm sw}\sim10^{-3}-10^{-2}\,\Msol$ in equipartition). As the blastwave is decelerating, the mass of the fastest [$(\Gamma\beta c)_{\rm eq}\sim0.55c$] ejecta responsible for the non-thermal emission is thus $M_{\rm ej}\sim0.01-0.1\,\Msol$ and has a kinetic energy of $\sim10^{51}-10^{52}$ erg. 

\subsection{Comparison to multi-wavelength FBOTs}\label{sec:FBOTs}

\sn{} and \cow{} are the only FBOTs for which we have long-term X-ray \emph{and} radio detections. ZTF18abvkwla is also detected at radio wavelengths \citep{Ho20}. Remarkably, the radio luminosity of the three FBOTs is large compared to SNe and some sub-energetic GRBs, and is even comparable to the radio emission in long GRBs (\koala{}).
Even with a sample of three radio-loud FBOTs, we already see a wide range of behaviors, which likely reflects a wide dynamic range of the properties of the fastest outflows of FBOTs. 

\koala{} and \sn{} share the presence of mildly relativistic, presumably jetted outflows (\S\ref{sec:nature_sn}). \koala{} had an expansion velocity\footnote{Equation (3) from \cite{Ho20} should read $\Theta=\frac{\Gamma\beta c t}{d_{A}(1+z)}$, leading to a revised $\Gamma \beta c>0.3c$ at $t_{obs}=81$ days (i.e. $\sim63$ days rest-frame). A. Ho, private communication.}\label{footnote:koala} (Fig. \ref{Fig:EnergyProfile}) of $\Gamma\beta c\ge 0.3c$ at $t\sim100$ days. They establish a class of transients that are able to launch relativistic ejecta with similarities to GRBs, yet differ from GRBs in their thermal optical emission (and presence of H, for \sn{}, Dong et al., in prep).
The mildly relativistic velocity of \sn{} and ZTF18abvkwla, and the large energy of the blast-wave in \sn{} differ distinctly from the non-relativistic and slow blast-wave in \cow{}, which showed $v\sim0.1c$ (Fig. \ref{Fig:EnergyProfile}, \citealt{Margutti19, Ho19}). Indeed, high spatial resolution radio observations of \cow{} indicated that \cow{} did not harbor a long-lived relativistic GRB-like jet \citep{Bietenholz20}.

The post-peak decline in radio luminosity of the radio-detected FBOTs is extraordinarily steep compared to all other classes of transients (Fig. \ref{Fig:RadioComparison}), even the energetic and highly collimated GRBs. \sn{} and \cow{} had comparable rates of $L_{8\,\rm{GHz}}\propto t^{-5.1\pm0.3}$ and $L_{8\,\rm{GHz}}\propto t^{-4.19\pm0.4}$ (Coppejans et al. in prep) respectively. The decline of ZTF18abvkwla \citep{Ho20} was shallower, with $L_{8\,\rm{GHz}}\propto t^{-2.7\pm0.4}$. A comparison between the radio properties of these three FBOTs also shows other spectral and evolutionary differences. Compared to \cow, which had $F_p\propto t^{-1.7\pm0.1}$ and $\nu_p\propto t^{-2.2\pm0.1}$ \citep{Margutti19,Ho19}, \sn{} exhibited a similar $F_p(t)$ evolution but a slower $\nu_p(t)$ decay. The information on the radio spectral properties of the FBOT ZTF18abvkwla is limited, but we note that at $\sim63$ days \cite{Ho20} infer $\nu_p\sim 10$ GHz with a significantly larger radio luminosity $L_{\nu}\sim 10^{30}\,\rm{erg\,s^{-1}}$ than \sn{} (Fig. \ref{Fig:RadioComparison}).

We now turn to the X-ray emission in \sn{} and \cow{}.
Although we only have late time X-ray observations of \sn{}, the luminosity appears to be consistent with that of \cow{} at $\sim100$ days post explosion (see Figure \ref{Fig:XrayComparison}). As was the case in \cow{}, the source of the X-ray emission cannot be synchrotron emission from the same population of electrons that produces the radio emission. In the two epochs at 99 and 357 days where we have simultaneous X-ray and radio observations, the extrapolated radio flux densities are consistent with the X-ray measurements only if we do not account for the presence of the synchrotron cooling break at $\nu=\nu_c$. For the $B_{eq}$ of Table \ref{Tab:properties}, we expect $\nu_c$ to lie between the radio and X-ray bands at $99<t<357$ days leading to a flux density steepening $F_{\nu}\propto \nu^{-p/2}\propto \nu^{-1.8}$ at $\nu>\nu_c$ \citep{Rybicki1979}. It follows that the extrapolated SSA spectrum under-predicts the X-ray flux and that another mechanism is thus required to explain the X-ray emission in \sn{}. In \cow{} there was also an excess of X-ray emission, which was attributed to a central engine \citep{Prentice18,Perley19,Kuin19,Margutti19,Ho19,Lyutikov19}. We speculate that the X-ray emission in \sn{} might also be attributable to the central engine. Interestingly, both FBOTs also have hydrogen-rich outflows (Dong et al.\ in prep.) and dense environments, and at optical/UV wavelengths are among the most luminous and fastest evolving members of the FBOT family (Dong et al., in prep.).

%%%%%%%%%%%%%%%%%%%%%%%%%%%%%%%%%%%%%%%%%%%%%%%%%%%%%%%%%%%%%%%%%%%%%%%%%%%%%%%%%%%%%%
\section{Properties of the dwarf host galaxy}
\label{Sec:host}

\begin{figure}
	\centering
	\includegraphics[width=0.47\textwidth]{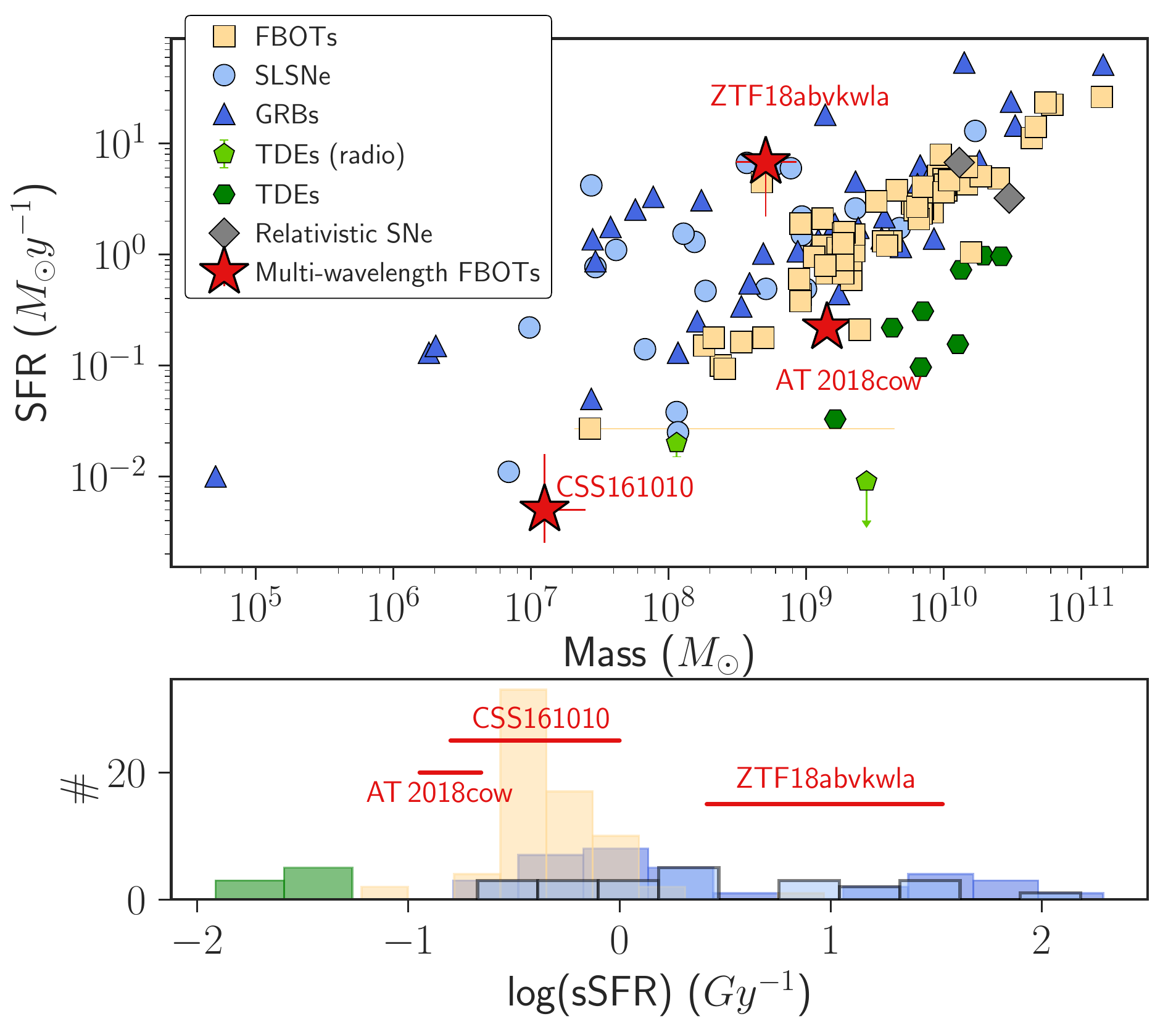}
    \caption{\emph{Upper panel}: Star Formation Rate  (SFR) and stellar mass properties of the dwarf galaxy host of \sn{} and other radio-loud FBOTs (red stars) in relation to the host galaxies of long GRBs (blue triangles), SLSNe (light-blue circles), FBOTs (squares) and TDEs (green diamonds). \emph{Lower panel:} histogram of specific SFR (sSFR) for the same classes of transients using the same color coding as above. The dwarf host of \sn{} has the lowest SFR of the FBOT hosts and a very small stellar mass $M_{\star}\sim10^{7}\,\rm{M_{\odot}}$. However, its sSFR is similar to those of other FBOTs, SLSNe and GRBs. References: \cow{} \citep{Perley19}, \koala{} \citep{Ho20}, TDEs (\citealt{Holoien16,Law-Smith2017,Saxton17} and private communication with Paulina Lira), FBOTs \citep{Drout14,Arcavi16,Pursiainen18}, SLSNe I \citep{Lunnan14}, relativistic SNe \citep{Michalowski18}, GRBs \citep{Svensson10}.}
    \label{Fig:host_SFR_mass}
\end{figure}

We use the Fitting and Assessment of Synthetic Templates code (FAST \citealt{Kriek09}) to fit the host galaxy emission and constrain the properties of the underlying stellar population. We first combine and re-normalize the Keck-LRIS spectrum by using the broad-band PanSTARSS \emph{gri} and DEIMOS \emph{VRI} photometry corrected for Galactic extinction. We do not include the NIR data at $\lambda\ge10000$ \AA\, (i.e., \emph{JHK} and the WISE \emph{W1} and \emph{W2} bands) in our fits, as these wavelengths are dominated by emission from the contaminating object (\S\ref{SubSec:host}). We assumed a \cite{Chabrier03} stellar initial mass function (IMF) and considered a variety of star formation histories and stellar population libraries. The best-fitting synthetic spectrum, which we show in Fig. \ref{Fig:host}, uses the stellar models of \cite{Bruzual03} with a metallicity of $Z=0.004$ and no internal extinction ($A_V=0$ mag), and favors an exponentially declining star formation law yielding a current star formation rate of $SFR\sim4\times 10^{-3}\,\Msol\,{\rm yr}^{-1}$. The total stellar mass of the host galaxy is $M_{*}\sim10^{7}\,\Msol$, which implies a current specific star formation rate $sSFR\sim 0.3\,\rm{Gyr^{-1}}$. Other choices of stellar population models, star formation histories and metallicity produce similar results. For example, using the stellar models of \cite{Bruzual03} and \cite{Conroy10}, with either an exponential or delayed exponential star formation history, and considering metallicity values in the range $Z=0.0008-0.02$ we find $A_V=0-0.4$ mag, a current stellar age of $(0.6-4)$ Gyr, a stellar mass of $M_{*}=(1-3)\times 10^{7}\,\Msol$, $SFR=(0.3-2)\times 10^{-2}\,\Msol\,{\rm yr^{-1}}$ and $sSFR=(0.2-1)\,\rm{Gyr^{-1}}$. The star formation rates that we derive using the [OII] and H$\alpha$ spectral lines are consistent with the value derived from our models.

Figure~\ref{Fig:host_SFR_mass} shows the properties of \sn's host compared to those of the hosts of other relevant classes of explosive transients. Interestingly, \sn{} has the smallest host mass of the known FBOTs, with the three radio-loud FBOTs known (red stars and symbols) populating the low-mass end of the host galaxy distribution. Hydrogen-stripped superluminous supernovae (SLSNe I) and long GRBs also show a general preference for low mass and low metallicity hosts (\S\ref{sec:nature_sn} for further discussion). 
It is important to note that the star formation rate per unit mass of the host of \sn{} is comparable to that of other transient classes involving massive stars.

We conclude this section by commenting that there is no observational evidence of activity from the dwarf host galaxy nucleus. There were no observed outbursts or flaring events (AGN-like activity) at the location of \sn{} prior to explosion. Specifically, we applied the Tractor image modeling code \citep{Lang16} across 6 g-band Dark Energy Camera  epochs (DECam, from 2018-10-06 to 2018-10-13) and 137 r-band and 3 g-band Palomar Transient Factory images (PTF, from 2009-10-03 to 2014-11-13) to find the best fit model for a host galaxy profile and a point source close to the position of \sn{}. We find no evidence for the presence of a variable point source in either DECam (\citealt{Dey19}) or PTF images prior to explosion of \sn{} (2016 October 6).

%%%%%%%%%%%%%%%%%%%%%%%%%%%%%%%%%%%%%%%%%%%
\section{The intrinsic nature of \sn{}}\label{sec:nature}

The key properties of \sn{} can be summarized as follows: it had a rise-time of a few days in the optical and showed a large peak optical luminosity of $\sim10^{44}$ erg~s$^{-1}$ (Dong et al.\ in prep.). Broad H$\alpha$ features also indicate that there was hydrogen in the outflow (Dong et al.\ in prep.). The surrounding CSM has a  
large density corresponding to an effective mass-loss of $\dot M\sim 2\times 10^{-4}\,\rm{M_{\sun}yr^{-1}}$ (for $v_w=1000\,\rm{km\,s^{-1}}$) at $r\sim10^{17}$ cm. The dwarf host galaxy has a stellar mass of $\sim 10^7\,\Msol$ that is significantly lower than other FBOTs, but has a comparable sSFR (Figure \ref{Fig:host_SFR_mass}). From our radio modelling, we know that the outflow was mildly relativistic with initial $\Gamma\beta c >$0.6c. The fast outflow has an ejecta mass of $\sim0.01-0.1\,\Msol$ and a kinetic energy of $E_k\gtrsim10^{51}$ erg. The X-ray emission is not produced by the same electrons producing the radio emission.

\begin{table*}
\centering
    \caption{Volumetric rate estimates for the entire population of FBOTs (upper part) and for the most luminous FBOTs (lower part).}
    \label{Tab:FBOTrate}
    \begin{tabular}{p{1.9cm}cccccccc}
    \hline
Reference &Abs Mag Range & Timescale & z & FBOT Rate & vs.  & vs. & vs.\\
& at peak (mag) &(days)$^{a}$ & & ($\rm{Gpc^{-3}yr^{-1}}$) & CCSNe$^{\rm{b}}$ &  SLSNe$^{\rm{c}}$ & sub-E GRBs$^{\rm{d}}$\\ 
\hline
\raggedright{\cite{Drout14}} & $-20 <\rm{M_{g}}<-16.5$ & $<12$&$<0.65$ & $4800-8000$ & 7-11\% & 2400-4000\% & 2100-3500\%\\
\raggedright{\cite{Pursiainen18}} &  $-15.8<\rm{M_{g}}<-22.2$ & $<10$& $0.05\le z \le 1.56$ & $\gtrsim 1000$ & $\gtrsim1.4\%$ & $\gtrsim500\%$ & $\gtrsim430\%$\\
\raggedright{\cite{Tampo20}} & $-17<\rm{M_{i}}<-20$ & $\lesssim 15$ & $0.3\le z\le 1.5$ & $\sim$4000 & $\sim$6\%& $\sim$2000\%&$\sim$1700\%\\
\hline
\raggedright{\cite{Ho20}} & $\rm{M_{g}}<-20$ & $<5$ & $\lesssim 0.1$ & $<$560 & $<0.8$\% &$<280$\% & $<240$\%\\
\raggedright{This work (PS1-MDS)} & $\rm{M_{g}}<-19$ & $<12$ &$<0.65$ & 700-1400 & 1-2\% & 350-700\% & 300-600\%\\
\raggedright{This work (PTF)} & $\rm{M_{R}}=-20 \pm 0.3$ & $\lesssim 3$ &$\lesssim 0.1$ & $<$300 &  $<0.4$\% & $<150\%$ & $<130\%$\\
\hline
\end{tabular}
\tablecomments{\footnotesize{$^{\rm{a}}$ Rest frame. $^{\rm{b}}$ Local Universe core-collapse SN rate from \cite{Li11} $\Re\sim70500\,\rm{Gpc^{-3}yr^{-1}}$ $^{\rm{c}}$ SLSN rate at $z\sim0.2$ from \cite{Quimby13}, including type I and type II events $\Re\sim200\,\rm{Gpc^{-3}yr^{-1}}$. $^{\rm{d}}$ Rate of sub-energetic long GRBs before beaming correction from \cite{Soderberg06c} $\Re\sim230\,\rm{Gpc^{-3}yr^{-1}}$. }}
\end{table*}

\subsection{Volumetric Rates of the most luminous FBOTs in the local Universe}\label{sec:rates}

We present three independent rate estimates for FBOTs such as \sn{}, \cow{} and ZTF18abvkwla, which populate the most luminous end of the optical luminosity distribution of FBOTs with optical bolometric peak luminosity $L_{opt} \gtrsim 10^{44}\,\rm{erg\,s^{-1}}$. At the end of this section we compare our estimates to the inferences by \cite{Ho20} and \cite{Tampo20}, which were published while this work was in an advanced stage of  preparation.

\citet{Drout14} determined an intrinsic rate for FBOTs with  absolute magnitude $-16.5 \ge M \ge -20$ of 4800$-$8000 events Gpc$^{-3}\rm{y}^{-1}$ based on the detection efficiency of the PanSTARRS1 Medium Deep Survey (PS1-MDS) for fast transients as a function of redshift. However, this estimate assumes a Gaussian luminosity function with a mean and variance consistent with the \emph{entire} PS1-MDS population of FBOTs, after correcting for detection volumes. In order to assess the intrinsic rate of \emph{luminous} rapid transients, such as \sn{}, we repeat the rate calculation of \citet{Drout14}, but adopt a new luminosity function based only on the four PS1-MDS events brighter than $-$19 mag in the $g$-band (PS1-11qr, PS1-12bbq, PS1-12bv, and PS1-13duy). This yields intrinsic rates for FBOTs with peak magnitudes greater than $-$19 mag of 700$-$1400 Gpc$^{-3}\rm{y}^{-1}$, which is $\sim$0.6$-$1.2\% of the core-collapse SN rate at $z\sim 0.2$ from \citet{Botticella2008} or $\sim1-2$\% of the local ($<60$ Mpc) core-collapse SN rate from \citet{Li11_LOSS}.

We further estimated the luminous FBOT rate from the Palomar Transient Factory \citep[PTF;][]{Law09PTF,Rau09PTF}. The PTF was an automated optical sky survey that operated from 2009--2012 across $\sim$8000\,deg$^2$, with cadences from one to five days, and primarily in the Mould $R$-band. We adopted the PTF detection efficiencies of \citet{Frohmaier17} and simulated a population of FBOTs with light curves identical to \cow{} (as we have color information for \cow{} near optical peak) and a 
Gaussian luminosity function M$_{R}=-20\pm0.3$ mag. Our methodology closely follows that described in \citet{Frohmaier18}, but with a simulation volume set to $z \le 0.1$ to maintain high completeness. We also performed a search for \cow{}-like events in the PTF data and found zero candidates. Given both the results of our simulations and no comparable events in the data, we measure a $3\sigma$ upper limit on the luminous FBOTs rate to be 300 Gpc$^{-3}\rm{y}^{-1}$, which is $\lesssim0.25\%$ of the core-collapse SN rate at $z \sim 0.2$ \citep{Botticella2008} or $\lesssim0.4\%$\ of the local core-collapse SN rate \citep{Li11_LOSS}. This volumetric rate is consistent with what we derive for luminous FBOTs in massive galaxies based on the Distance Less Than 40 Mpc survey (DLT40, \citealt{Tartaglia18}) following \citealt{Yang17}. We refer to the PTF rate estimate in the rest of this work.

We compare our rate estimates of luminous FBOTs in the local Universe ($z\le 0.1$) with those derived by \cite{Ho20} from the archival search of 18 months of ZTF-1DC survey. The transient selection criteria by \cite{Ho20} are comparable to our set up of the simulations on the PTF data set. Specifically, \cite{Ho20} selected transients with peak absolute $g$-band magnitude $M_{g,pk}<-20$ and rapid rise time $<5$ days, finding a limiting volumetric rate $<400\,\rm{Gpc^{-3}yr^{-1}}$ at distances $<560$ Mpc, consistent with our inferences.  Our study and \cite{Ho20} thus independently identify luminous FBOTs as an intrinsically rare type of transient, with a volumetric rate $<(0.4-0.6)\%$ the core-collapse SN rate in the local Universe. We conclude that luminous FBOTs are sampling a very rare channel of stellar explosion or other rare phenomenon (\S\ref{sec:models}). Interestingly the luminous FBOT rate is potentially comparable to that of sub-energetic long GRBs ($230^{+490}_{-190}\,\rm{Gpc^{-3}yr^{-1}}$, 90\% c.l., before beaming correction, \citealt{Soderberg06c}), and local SLSNe ($199^{+137}_{-86}\,\rm{Gpc^{-3}yr^{-1}}$ at $z=0.16$, \citealt{Quimby13}).  

We end by noting that our rate estimates are \emph{not} directly comparable to those inferred by \cite{Tampo20} from the HSC-SSP transient survey. These authors considered rapidly evolving transients in a wider range of luminosities ($-17\ge M_{i}\ge -20$) at cosmological distances corresponding to $0.3\le z\le1.5$ and inferred a rate $\sim4000\,\rm{Gpc^{-3}yr^{-1}}$. A similar argument applies to the FBOT rates by \cite{Pursiainen18}. Table \ref{Tab:FBOTrate} presents a summary of the current estimates of the volumetric rate for both the entire population of FBOTs and for the most luminous FBOTs.

%%%%%%%%%%%%%%%%%%%%%%%%%%%%%%%%%%%%%%%%%%%
\subsection{Physical Models}\label{sec:models}
Multiple physical models have been suggested to explain the optical behaviour of FBOTs (see \S\ref{Sec:intro}). Here, we consider mechanisms/transients that could power the radio and X-ray emission of the FBOT \sn{}. As the ejecta is hydrogen-rich (Dong et al., in prep.), we do not consider neutron star mergers and accretion induced collapse models. We also disfavour models involving the disruption or explosion of white dwarfs (WDs). 

\sn{} is not flaring activity associated with an Active Galactic Nucleus (AGN). The fraction of dwarf galaxies with masses of the order $10^7\,\Msol$ that host an AGN is not well-constrained \citep[e.g.,][]{Mezcua18}, but as there is at least one AGN host with a stellar mass comparable to \sn{} ($1-3\times10^7 \, \Msol$ \citealt{Mezcua18}), an AGN cannot be excluded based on the small host galaxy mass alone. The evolving synchrotron radio spectrum is not consistent with the typical flat spectrum seen in AGNs. There is also no evidence for prior optical or radio variability in PTF data (\S\ref{Sec:host}) or the NRAO/VLA Sky Survey (NVSS, \citealt{Condon98}). Most importantly, the optical line flux ratios of [\ion{N}{2}]$\lambda$6584/H$\alpha$ vs [\ion{O}{3}]/$\lambda$5007/H$\beta$ \citep{Kewley02,Kauffmann03} from our Keck spectrum (Fig. \ref{Fig:host}) exclude the presence of an AGN.

%%%%%%%%%%%%%%%%%%%%%%%
\subsubsection{Stellar Explosion}\label{sec:nature_sn}
In \S\ref{sec:mildly_rel} we inferred that \sn{} has $E_k>6\times10^{49}\,\rm{erg}$ coupled to fast moving material with $\Gamma\beta c \ge0.55c$. This finding implies that the slow moving material at $v\sim10,000\,\rm{km\,s^{-1}}$ would have $E_k>10^{53}\,\rm{erg}$ under the standard scenario of a spherical hydro-dynamical collapse of a star, where $E_k\propto (\Gamma\beta)^{-\alpha}$ with $\alpha\approx-5.2$ for a polytropic index of 3 \citep{Tan01}. This value largely exceeds the $E_{k}\sim10^{51}$ erg limit typical of neutrino-powered stellar explosions, pointing to a clear deviation from a spherical collapse. We conclude that if \sn{} is a stellar explosion, then its fastest outflow component (i.e. the one powering the radio emission that we detected at late times) must have been initially aspherical and potentially jetted, similar to that of GRBs.  Indeed, Fig. \ref{Fig:EnergyProfile} shows that only GRBs (and jetted TDEs) have comparable energy coupled to their relativistic outflows, suggesting that regardless of the exact nature of \sn{}, a compact object (such as a magnetar or accreting black hole) is necessary to explain the energetics of its outflow. In the context of SNe, \sn{} thus qualifies as an engine-driven explosion. 

This finding has important implications. Shock interaction with, or breakout from, a dense confined shell of material surrounding the progenitor has been proposed to explain the blue optical colors and fast optical evolution of a number of FBOTs \citep[e.g.,][]{Drout14,Whitesides17}. Although these mechanisms could explain the optical colors and fast rise times of FBOTs, they cannot naturally produce the mildly relativistic outflows observed in \sn{} (and ZTFabvkwala, \citealt{Ho20}). We thus conclude that a pure shock interaction/breakout scenario of a normal SN shock through a dense medium cannot account for all the properties of luminous FBOTs across the electromagnetic spectrum, and that at least some luminous FBOTs are \emph{also} powered by a central engine, as it was inferred for \cow{} \citep{Margutti19,Ho19,Perley19}. The analysis of ZTF18abvkwla by \cite{Ho20} supports a similar conclusion.

Known classes of engine-driven stellar explosions include relativistic SNe, (long) GRBs, and SLSNe. The dwarf nature of the host galaxies of luminous FBOTs that are engine-driven (red stars in Fig. \ref{Fig:host_SFR_mass}) is reminiscent of that of some SLSNe and GRBs, which show a preference for low-mass galaxies \citep[e.g.,][]{Lunnan14,Chen17,Schulze18}, as independently pointed out by \cite{Ho20}.
A second clear similarity between luminous FBOTs, relativistic SNe and GRBs is the presence of relativistic outflows (Fig. \ref{Fig:EnergyProfile}) and the associated luminous radio emission (Fig. \ref{Fig:RadioComparison}), which is clearly not present with similar luminosities in SLSNe (\citealt{Coppejans18,Eftekhari19,Law19}).\footnote{There is only one SLSN detected to date \citep{Eftekhari19,Law19} out of the few dozen observed at radio wavelengths \citep[e.g.,][ and references therein]{Coppejans18,Law19}. No jet has been detected in a SLSN  and for H-stripped SLSNe the radio limits rule out off-axis jets in the lower energy and density range of GRBs \citep{Coppejans18}.}
Yet, luminous FBOTs differ from any known class of stellar explosions with relativistic ejecta in two key aspects: (i) the temporal evolution and spectroscopic properties of their thermal UV/optical emission; (ii) \sn{} showed evidence for a large mass coupled to its fastest (relativistic) outflow. We expand on these major differences below.

Luminous FBOTs with multi-wavelength detections reach optical bolometric peak luminosities $\gtrsim 10^{44}\,\rm{erg\,s^{-1}}$  (\citealt{Prentice18,Perley19, Margutti19}; Dong et al.\ in prep.) comparable only to SLSNe. The extremely fast temporal evolution (over time-scales of $\sim$days) and hot, mostly featureless initial spectra with $T\sim40,000\,K$ (\citealt{Kuin19, Perley19,Margutti19, Ho20}) distinguish luminous FBOTs from any other engine-driven transients. While it is unclear if the ejecta of ZTF18abvkwla contained hydrogen \citep{Ho20}, \cow{} and \sn{} showed for hydrogen rich ejecta (\citealt{Margutti19, Prentice18,Perley19}, Dong et al.\ in prep.). In fact, \sn{} is the first case where a relativistic hydrogen-rich outflow is observed, which implies the existence of a new class of engine-driven explosion that originate from progenitors that still retain a significant fraction of their hydrogen envelope at the time of explosion. There are some reasons to expect that jets should be preferentially launched in explosions of hydrogen-stripped progenitors. For example, binary interaction can strip away the stellar envelope while spinning up the core. The angular momentum of the core is an important ingredient for launching jets. Alternatively jets in hydrogen-rich progenitors could simply lack the necessary energy to pierce through the stellar envelope. \citep[e.g.,][and references therein]{MacFadyen99,MacFadyen01,Lazzati12,Bromberg11,Nakar12,Margutti14b}.  

Next we comment on the amount of mass coupled to the fastest ejecta. While the shock velocity of \sn{} is comparable to that of the relativistic SNe and the initial $E_k$ of the outflow is similar to GRBs, the fastest ejecta mass of \sn{} is significantly larger than that of GRB jet outflows, which typically carry $\sim10^{-6}-10^{-5}\,\Msol$. It thus comes as no surprise that  neither on- nor off-axis GRB-like jet models \citep[e.g.,][]{Granot02,vanEerten2012} fit the radio temporal or spectral evolution of \sn{}. Indeed, the ejecta mass carried by GRB jets needs to be small enough to reach sufficiently large velocities to prevent the absorption of $\gamma$-rays for pair production \citep[see][]{Dermer2000,Huang02,Nakar03}. Explosions with a sufficiently large ejecta mass to be important in the dynamics and absorb the high-energy emission are referred to as `baryon-loaded explosions' or `dirty fireballs'. Although predicted \citep[e.g.,][]{Huang02}, such sources have remained fairly elusive. The relativistic SN\,2009bb is argued to be relativistic and baryon-loaded with $M_{\rm ej}\geq10^{-2.5}\,\Msol$ \citep{Chakraborti11}, and the transient PTF11agg is another potential relativistic baryon-loaded candidate \citep{Cenko13}. \sn{} is mildly relativistic, did not have a detected gamma-ray counterpart (\S\ref{sec:gammarays}), had a large $E_k$ that is comparable to GRBs, and had an ejecta mass that is intermediate between GRBs and SNe. It is thus a relativistic baryon-loaded explosion or dirty fireball. Interestingly, luminous GRB-like $\gamma$-ray emission was also ruled out for the other mildly relativistic FBOT ZTF18abvkwla \citep{Ho20}.

Our major conclusion is that while luminous multi-wavelength FBOTs share similarities with other classes of engine driven explosions, their properties clearly set them apart as a completely new class of engine-driven transients comprising at most a very small fraction of stellar deaths (\S\ref{sec:rates}). Special circumstances are thus needed to create the most luminous FBOTs.

%%%%%%%%%%%%%%%%%%%%%
\subsubsection{Tidal Disruption Event by an Intermediate Mass Black Hole}

One of the proposed models for the FBOT \cow{} was a tidal disruption event (TDE) of a star by an intermediate mass black hole (IMBH, \citealt{Perley19,Kuin19}). \citet{Margutti19} disfavour this model as it is difficult to explain the origin of the high-density surrounding medium (inferred from radio observations) with a TDE on an off-center IMBH. \sn{} is spatially consistent with the nucleus of its host, so this argument is not directly applicable here.

The dwarf host galaxy of \sn{} is at least $\sim10$ times less massive than any other confirmed TDE host (Fig. \ref{Fig:host}). The $M_{\star}\approx10^7\,\Msol$ implies that the central BH would likely be an IMBH. The BH masses and occupation fractions in dwarf galaxies are not well constrained. However, using the relations between the BH mass and host galaxy stellar mass in \citet{Marleau13} and \citet{Reines15}, which were derived largely based on higher mass galaxies, we obtain a rough estimate for the BH mass of $\sim10^3\,\Msol$. For this BH mass, the X-ray luminosity at $\sim100$ days is $\sim0.01\,L_{\rm Edd}$ (where $L_{\rm Edd}$ is the Eddington Luminosity) and the optical bolometric luminosity is $10^3\,L_{\rm Edd}$. The optical luminosity would have to be highly super-Eddington in this scenario. However, the optical luminosity estimate is highly dependent on the assumed temperature, the uncertainty on the BH mass is very large, and \sn{} was aspherical and clearly showed an outflow. Consequently we cannot conclusively rule out that \sn{} is a TDE based on the luminosity. 

It is similarly not possible to rule out a TDE scenario based on the optical rise and decay time-scales. It is true that the optical rise and decay rate of \sn{} was significantly faster than TDEs on super-massive black-holes, SMBHs \citep[e.g.][]{Hinkle20}. In fact, the $\sim4$ day optical rise of \sn{} (Dong et al., in prep.) was shorter than the $\sim11$ day rise of the fastest TDE discovered to date iPTF16fnl (which had a BH mass of $\leq10^{6.6}\,\Msol{}$ \citealt{Blagorodnova17}) and formally consistent with the classical TDE scalings $t_{rise}\sim 1.3 (M_{\rm{BH}}/10^{3}\,\rm{M_{\sun}})^{1/2}$ days for a Sun-like star disruption.
However, the circularization of the debris is unlikely to be efficient and the circularization timescales of the debris are highly uncertain for IMBHs \citep[e.g.,][]{Chen18,Darbha19} and we cannot directly compare the TDE timescales of SMBHs and IMBHs. 
The radio and X-ray luminosities of \sn{} are comparable to those of some jetted TDEs (Figures \ref{Fig:RadioComparison} and \ref{Fig:XrayComparison}), although \sn{} shows a faster radio decline. The kinetic energy is also comparable to the jetted TDEs (Figure \ref{Fig:EnergyProfile}). In TDEs that lack gamma-ray detections, the radio synchrotron emission is proposed to be from the shock between the CSM and an outflow driven by a super-Eddington accretion phase \citep[e.g.,][]{Rees88,Strubbe09,Zauderer11,Alexander16}, or
the external shock from the unbound stellar material \citep{Krolik16}, or internal shocks in a freely expanding relativistic jet \citep{Pasham18}. The outflows are modelled using equipartition analysis as we have done for \sn{} in Section \ref{Sec:Radio_Xray_inferences}, so our results are equally applicable to TDE models and we cannot rule out a TDE based on the radio properties.

Based on the aforementioned arguments, and the fact that the dwarf host galaxy spectrum does not have clear post-starburst features, we disfavour the scenario that \sn{} is a TDE of an IMBH but cannot conclusively exclude it. If this scenario is true though, then there are several implications. First, as \sn{} is hydrogen rich (Dong et al.\ in prep.), the disrupted star would likely not be a WD. Second, \sn{} would be the TDE with the smallest BH mass to date. This would imply that TDEs on IMBHs can produce transients that launch relativistic outflows and show short rise times of a few days. If this is the case, then multi-wavelength observations of FBOTs could identify IMBHs and also help to determine the BH mass function and occupation fraction at low galaxy masses. Third, the volumetric rates estimates for SMBH TDEs are $\sim200$ Gpc$^{-3}\rm{y}^{-1}$ (Alexander et al.\ submitted). If the population of luminous FBOTs is the population of TDEs on IMBHs, then our volumetric rate estimate for luminous FBOTs ($\lesssim300$ Gpc$^{-3}\rm{y}^{-1}$) would imply that the rate of TDEs on IMBHs would be at most that of the TDE rate of SMBHs.

%%%%%%%%%%%%%%%%%%%%%%%%%%%%%%%%%%%%%%%%%%%
\section{Summary and Conclusions}
\label{Sec:Conc}
We present X-ray and radio observations of the luminous FBOT \sn{} and its dwarf host galaxy. The optical properties of the transient are described in Dong et al.\ in prep. At the distance of $\sim$150 Mpc, \sn{} is the second closest FBOT (after \cow{}). To date, \sn{} is one of only two FBOTs detected at radio \textit{and} X-ray wavelengths (including \cow{}, \citealt{RiveraSandoval18,Margutti19,Ho19}) and three detected at radio wavelengths (including \cow{} and \koala{}, \citealt{Ho20}). 

We highlight below our major observational findings:
\begin{itemize}
\item \sn{} reached a radio luminosity $L_{\nu}\sim 10^{29}$ erg~s$^{-1}$~Hz$^{-1}$ (at $\nu=6$ GHz) comparable to sub-energetic GRBs (i.e. significantly larger than normal SNe), with a steep after-peak temporal decline similar to that observed in \cow{}.
\item The radio properties of \sn{} imply the presence of a decelerating mildly relativistic outflow with $\Gamma\beta c>0.6c$ at $t$$=$$99$ days, carrying a large ejecta mass $\gtrsim 0.01\,\rm{M_{\odot}}$ and kinetic energy $E_k$$>$$10^{50}$ erg, and propagating into a dense environment with $n\approx700\,\rm{cm^{-3}}$ at $r\approx10^{17}$ cm (an effective mass-loss rate of $\dot{M}\approx2\times10^{-4}\,\Msol\,y^{-1}$ for a wind velocity of 1000 $\rm{km\,s^{-1}}$).
\item The X-ray luminosity of $3\times 10^{39}\,\rm{erg\,s^{-1}}$ (at 99 days) is too bright to be synchrotron emission from the same population of electrons powering the radio emission. In \cow{} this X-ray excess was attributed to a central engine and we speculate that this is also the case in \sn{}.
\item \sn{} resides in a small dwarf galaxy with  stellar mass $M_{*}$$\sim$$ 10^{7}\,\Msol$, (the smallest host galaxy to an FBOT to date). However, its specific star formation rate of $sSFR=(0.2-1)\,\rm{Gyr^{-1}}$ is comparable to other transient host galaxies (e.g. GRBs and SLSNe). Intriguingly, all the FBOTs with multi-wavelength detections so far have dwarf host galaxies (\citealt{Prentice18,Perley19,Ho20}).
\item \sn{}, \cow{} and \koala{} belong to a rare population of luminous FBOTs ($M_R<-20$ mag at peak). For this population, using PTF data, we estimate a volumetric rate $<$300 Gpc$^{-3}\rm{y}^{-1}$, which is $\lesssim0.4\%$ of the local core-collapse SN rate. This result is consistent with the estimates by \citet{Ho20}. We thus reach the same conclusion as \citet{Ho20} that luminous FBOTs stem from a rare progenitor pathway. 
\end{itemize}

In the context of stellar explosions, the properties of \sn{} imply a clear deviation from spherical symmetry (as in the case of GRB jets), and hence the presence of a ``central engine'' (black hole or neutron star) driving a potentially collimated mildly relativistic outflow. Differently from GRBs, \sn{} (i) has a significantly larger mass coupled to the relativistic outflow, which is consistent with the lack of detected $\gamma$-rays; (ii) the ejecta is hydrogen-rich (Dong et al., in prep.).  For \sn{} we cannot rule out the scenario of a stellar tidal disruption on an IMBH. However we  note that this scenario would imply a highly super-Eddington accretion rate of $\sim10^3\,L_{edd}$ for our (uncertain) BH mass estimate $\sim10^3\,\Msol{}$. Irrespective of the exact nature of \sn{}, \sn{}  establishes a new class of hydrogen-rich, mildly relativistic transients.

We end with a final consideration. The three known FBOTs that are detected at radio wavelengths are among the most luminous and fastest-rising among FBOTs in the optical regime (\citealt{Perley19,Margutti19,Ho19,Ho20}, Dong et al., in prep.).  Intriguingly, all the multi-wavelength FBOTs also have evidence for a compact object powering their emission \citep[e.g.,][]{Prentice18,Perley19,Kuin19,Margutti19,Ho19}.
We consequently conclude, independently of (but consistently with) \citealt{Ho20}, that at least some luminous FBOTs must be engine-driven and cannot be accounted for by existing FBOT models that do not invoke compact objects to power their emission across the electromagnetic spectrum.  Furthermore, even within this sample of three luminous FBOTs with multiwavelength observations, we see a wide diversity of properties of their fastest ejecta. While \sn{} and  \koala{} harbored mildly relativistic outflows, \cow{} is instead non-relativistic. Radio and X-ray observations are critical to understanding the physics of this intrinsically rare and diverse class of transients.

%%%%%%%%%%%%%%%%%%%%%%%%%%%%%%%%%%%%%%%%%%%
\bigskip
\bigskip\bigskip\bigskip
\section*{Acknowledgments}
We thank the anonymous referee for their feedback. We also thank the entire Chandra team, the VLA and the GMRT for their work, time and dedication that made these observations possible. We also thank Subo Dong, Brian Metzger, Kris Stanek, Cliff Johnson, Rocco Coppejans, Todd Thompson and Giacomo Fragione for useful discussions and comments.
%Grants:
The Margutti's team at Northwestern is partially funded by the Heising-Simons Foundation under grant \# 2018-0911 (PI: Margutti) and by NASA Grant \#80NSSC19K0384. R.M. acknowledges support by the National Science Foundation under Award No. AST-1909796. Raffaella Margutti is a CIFAR Azrieli Global Scholar in the Gravity \& the Extreme Universe Program, 2019. This research was supported in part by the National Science Foundation under Grant No. NSF PHY-1748958.
K.D.A. and E.R.C. acknowledge support provided by NASA through the Hubble Fellowship Program, grants HST-HF2-51403.001 and HST-HF2-51433.001-A, awarded by the Space Telescope Science Institute, which is operated by the Association of Universities for Research in Astronomy, Inc., for NASA, under contract NAS5-26555. CSK is supported by NSF grants AST-1908952 and AST-1814440. Research by DJS is supported by NSF grants AST-1821967, 1821987, 1813708, 1813466, and 1908972. TAT is supported in part by NASA Grant \#80NSSC18K0526. B.J.S. is supported by NSF grants AST-1908952, AST-1920392, and AST-1911074. M.R.D. acknowledges support from the Dunlap Institute at the University of Toronto and the Canadian Institute for Advanced Research (CIFAR). BAZ acknowledges support from the DARK Cosmology Centre and while serving at the National Science Foundation. D.S., D.F., and A.R. gratefully acknowledge support from RFBR grant 18-02-00062. 
%VLA:
The National Radio Astronomy Observatory is a facility of the National Science Foundation operated under cooperative agreement by Associated Universities, Inc. GMRT is run by the National Centre for Radio Astrophysics of the Tata Institute of Fundamental Research.
%Chandra:
The scientific results reported in this article are based in part on observations made by the Chandra X-ray Observatory. This research has made use of software provided by the Chandra X-ray Center (CXC) in the application packages CIAO.
%Keck:
The data presented herein were obtained at the W. M. Keck Observatory, which is operated as a scientific partnership among the California Institute of Technology, the University of California and the National Aeronautics and Space Administration. The Observatory was made possible by the generous financial support of the W. M. Keck Foundation.
The authors wish to recognize and acknowledge the very significant cultural role and reverence that the summit of Maunakea has always had within the indigenous Hawaiian community. We are most fortunate to have the opportunity to conduct observations from this mountain.
%Pan-STARRS1:
The Pan-STARRS1 Surveys (PS1) and the PS1 public science archive have been made possible through contributions by the Institute for Astronomy, the University of Hawaii, the Pan-STARRS Project Office, the Max-Planck Society and its participating institutes, the Max Planck Institute for Astronomy, Heidelberg and the Max Planck Institute for Extraterrestrial Physics, Garching, The Johns Hopkins University, Durham University, the University of Edinburgh, the Queen's University Belfast, the Harvard-Smithsonian Center for Astrophysics, the Las Cumbres Observatory Global Telescope Network Incorporated, the National Central University of Taiwan, the Space Telescope Science Institute, the National Aeronautics and Space Administration under Grant No. NNX08AR22G issued through the Planetary Science Division of the NASA Science Mission Directorate, the National Science Foundation Grant No. AST-1238877, the University of Maryland, Eotvos Lorand University (ELTE), the Los Alamos National Laboratory, and the Gordon and Betty Moore Foundation.
%Wise:
This publication makes use of data products from the Wide-field Infrared Survey Explorer, which is a joint project of the University of California, Los Angeles, and the Jet Propulsion Laboratory/California Institute of Technology, funded by the National Aeronautics and Space Administration.
%CSS:
The CSS survey is funded by the National Aeronautics and Space Administration under Grant No. NNG05GF22G issued through the Science Mission Directorate Near-Earth Objects Observations Program. The CRTS survey is supported by the U.S.~National Science Foundation under grants AST-0909182 and AST-1313422. Research by DJS is supported by NSF grants AST-1821987 and 1821967. Research by SV is supported by NSF grants AST-1813176. IC's research is supported by the Smithsonian Astrophysical Observatory Telescope Data Center, the Russian Science Foundation grant 19-12-00281 and the Program of development at M.V. Lomonosov Moscow State University through the Leading Scientific School ``Physics of stars, relativistic objects and galaxies''. This publication makes use of the Interplanetary Network Master Burst List at ssl.berkeley.edu/ipn3/masterli.html.
%%%%%%%%%%%%%%%%%%%%%%%%%%%%%%%%%%%%%%%%%%%
\bibliographystyle{aasjournal}
\bibliography{master_sne}

\begin{thebibliography}{}
\expandafter\ifx\csname natexlab\endcsname\relax\def\natexlab#1{#1}\fi
\providecommand{\url}[1]{\href{#1}{#1}}
\providecommand{\dodoi}[1]{doi:~\href{http://doi.org/#1}{\nolinkurl{#1}}}
\providecommand{\doeprint}[1]{\href{http://ascl.net/#1}{\nolinkurl{http://ascl.net/#1}}}
\providecommand{\doarXiv}[1]{\href{https://arxiv.org/abs/#1}{\nolinkurl{https://arxiv.org/abs/#1}}}

\bibitem[{{Alexander} {et~al.}(2016){Alexander}, {Berger}, {Guillochon},
  {Zauderer}, \& {Williams}}]{Alexander16}
{Alexander}, K.~D., {Berger}, E., {Guillochon}, J., {Zauderer}, B.~A., \&
  {Williams}, P.~K.~G. 2016, \apj, 819, L25,
  \dodoi{10.3847/2041-8205/819/2/L25}

\bibitem[{{Alexander} {et~al.}(2017){Alexander}, {Wieringa}, {Berger},
  {Saxton}, \& {Komossa}}]{Alexander17}
{Alexander}, K.~D., {Wieringa}, M.~H., {Berger}, E., {Saxton}, R.~D., \&
  {Komossa}, S. 2017, \apj, 837, 153, \dodoi{10.3847/1538-4357/aa6192}

\bibitem[{{Arcavi} {et~al.}(2016){Arcavi}, {Wolf}, {Howell}, {Bildsten},
  {Leloudas}, {Hardin}, {Prajs}, {Perley}, {Svirski}, {Gal-Yam}, {Katz},
  {McCully}, {Cenko}, {Lidman}, {Sullivan}, {Valenti}, {Astier}, {Balland},
  {Carlberg}, {Conley}, {Fouchez}, {Guy}, {Pain}, {Palanque-Delabrouille},
  {Perrett}, {Pritchet}, {Regnault}, {Rich}, \& {Ruhlmann-Kleider}}]{Arcavi16}
{Arcavi}, I., {Wolf}, W.~M., {Howell}, D.~A., {et~al.} 2016, \apj, 819, 35,
  \dodoi{10.3847/0004-637X/819/1/35}

\bibitem[{{Berger} {et~al.}(2012){Berger}, {Zauderer}, {Pooley}, {Soderberg},
  {Sari}, {Brunthaler}, \& {Bietenholz}}]{Berger12}
{Berger}, E., {Zauderer}, A., {Pooley}, G.~G., {et~al.} 2012, \apj, 748, 36,
  \dodoi{10.1088/0004-637X/748/1/36}

\bibitem[{{Bertin} \& {Arnouts}(1996)}]{Bertin1996}
{Bertin}, E., \& {Arnouts}, S. 1996, \aaps, 117, 393,
  \dodoi{10.1051/aas:1996164}

\bibitem[{{Bietenholz} {et~al.}(2020){Bietenholz}, {Margutti}, {Coppejans},
  {Alexander}, {Argo}, {Bartel}, {Eftekhari}, {Milisavljevic}, {Terreran}, \&
  {Berger}}]{Bietenholz20}
{Bietenholz}, M.~F., {Margutti}, R., {Coppejans}, D., {et~al.} 2020, \mnras,
  491, 4735, \dodoi{10.1093/mnras/stz3249}

\bibitem[{{Blagorodnova} {et~al.}(2017){Blagorodnova}, {Gezari}, {Hung},
  {Kulkarni}, {Cenko}, {Pasham}, {Yan}, {Arcavi}, {Ben-Ami}, {Bue}, {Cantwell},
  {Cao}, {Castro-Tirado}, {Fender}, {Fremling}, {Gal-Yam}, {Ho}, {Horesh},
  {Hosseinzadeh}, {Kasliwal}, {Kong}, {Laher}, {Leloudas}, {Lunnan}, {Masci},
  {Mooley}, {Neill}, {Nugent}, {Powell}, {Valeev}, {Vreeswijk}, {Walters}, \&
  {Wozniak}}]{Blagorodnova17}
{Blagorodnova}, N., {Gezari}, S., {Hung}, T., {et~al.} 2017, \apj, 844, 46,
  \dodoi{10.3847/1538-4357/aa7579}

\bibitem[{{Blandford} \& {McKee}(1976)}]{Blandford76}
{Blandford}, R.~D., \& {McKee}, C.~F. 1976, Physics of Fluids, 19, 1130,
  \dodoi{10.1063/1.861619}

\bibitem[{{Botticella} {et~al.}(2008){Botticella}, {Riello}, {Cappellaro},
  {Benetti}, {Altavilla}, {Pastorello}, {Turatto}, {Greggio}, {Patat},
  {Valenti}, {Zampieri}, {Harutyunyan}, {Pignata}, \&
  {Taubenberger}}]{Botticella2008}
{Botticella}, M.~T., {Riello}, M., {Cappellaro}, E., {et~al.} 2008, \aap, 479,
  49, \dodoi{10.1051/0004-6361:20078011}

\bibitem[{{Bromberg} {et~al.}(2011){Bromberg}, {Nakar}, \&
  {Piran}}]{Bromberg11}
{Bromberg}, O., {Nakar}, E., \& {Piran}, T. 2011, \apjl, 739, L55,
  \dodoi{10.1088/2041-8205/739/2/L55}

\bibitem[{{Brown} {et~al.}(2017){Brown}, {Levan}, {Stanway}, {Kr{\"u}hler},
  {Tanvir}, {Davies}, {Fruchter}, {Cenko}, \& {Metzger}}]{Brown17}
{Brown}, G.~C., {Levan}, A.~J., {Stanway}, E.~R., {et~al.} 2017, \mnras, 472,
  4469, \dodoi{10.1093/mnras/stx2193}

\bibitem[{{Bruzual} \& {Charlot}(2003)}]{Bruzual03}
{Bruzual}, G., \& {Charlot}, S. 2003, \mnras, 344, 1000,
  \dodoi{10.1046/j.1365-8711.2003.06897.x}

\bibitem[{{Cenko} {et~al.}(2012{\natexlab{a}}){Cenko}, {Bloom}, {Kulkarni},
  {Strubbe}, {Miller}, {Butler}, {Quimby}, {Gal-Yam}, {Ofek}, {Quataert},
  {Bildsten}, {Poznanski}, {Perley}, {Morgan}, {Filippenko}, {Frail}, {Arcavi},
  {Ben-Ami}, {Cucchiara}, {Fassnacht}, {Green}, {Hook}, {Howell}, {Lagattuta},
  {Law}, {Kasliwal}, {Nugent}, {Silverman}, {Sullivan}, {Tendulkar}, \&
  {Yaron}}]{Cenko2012_PTF10iya}
{Cenko}, S.~B., {Bloom}, J.~S., {Kulkarni}, S.~R., {et~al.} 2012{\natexlab{a}},
  \mnras, 420, 2684, \dodoi{10.1111/j.1365-2966.2011.20240.x}

\bibitem[{{Cenko} {et~al.}(2012{\natexlab{b}}){Cenko}, {Krimm}, {Horesh},
  {Rau}, {Frail}, {Kennea}, {Levan}, {Holland}, {Butler}, {Quimby}, {Bloom},
  {Filippenko}, {Gal-Yam}, {Greiner}, {Kulkarni}, {Ofek}, {Olivares E.},
  {Schady}, {Silverman}, {Tanvir}, \& {Xu}}]{Cenko12}
{Cenko}, S.~B., {Krimm}, H.~A., {Horesh}, A., {et~al.} 2012{\natexlab{b}},
  \apj, 753, 77, \dodoi{10.1088/0004-637X/753/1/77}

\bibitem[{{Cenko} {et~al.}(2013){Cenko}, {Kulkarni}, {Horesh}, {Corsi}, {Fox},
  {Carpenter}, {Frail}, {Nugent}, {Perley}, {Gruber}, {Gal-Yam}, {Groot},
  {Hallinan}, {Ofek}, {Rau}, {MacLeod}, {Miller}, {Bloom}, {Filippenko},
  {Kasliwal}, {Law}, {Morgan}, {Polishook}, {Poznanski}, {Quimby}, {Sesar},
  {Shen}, {Silverman}, \& {Sternberg}}]{Cenko13}
{Cenko}, S.~B., {Kulkarni}, S.~R., {Horesh}, A., {et~al.} 2013, \apj, 769, 130,
  \dodoi{10.1088/0004-637X/769/2/130}

\bibitem[{{Chabrier}(2003)}]{Chabrier03}
{Chabrier}, G. 2003, \pasp, 115, 763, \dodoi{10.1086/376392}

\bibitem[{{Chakraborti} \& {Ray}(2011)}]{Chakraborti11}
{Chakraborti}, S., \& {Ray}, A. 2011, \apj, 729, 57,
  \dodoi{10.1088/0004-637X/729/1/57}

\bibitem[{{Chakraborti} {et~al.}(2015){Chakraborti}, {Soderberg}, {Chomiuk},
  {Kamble}, {Yadav}, {Ray}, {Hurley}, {Margutti}, {Milisavljevic},
  {Bietenholz}, {Brunthaler}, {Pignata}, {Pian}, {Mazzali}, {Fransson},
  {Bartel}, {Hamuy}, {Levesque}, {MacFadyen}, {Dittmann}, {Krauss}, {Briggs},
  {Connaughton}, {Yamaoka}, {Takahashi}, {Ohno}, {Fukazawa}, {Tashiro},
  {Terada}, {Murakami}, {Goldsten}, {Barthelmy}, {Gehrels}, {Cummings},
  {Krimm}, {Palmer}, {Golenetskii}, {Aptekar}, {Frederiks}, {Svinkin}, {Cline},
  {Mitrofanov}, {Golovin}, {Litvak}, {Sanin}, {Boynton}, {Fellows}, {Harshman},
  {Enos}, {von Kienlin}, {Rau}, {Zhang}, \& {Savchenko}}]{Chakraborti15}
{Chakraborti}, S., {Soderberg}, A., {Chomiuk}, L., {et~al.} 2015, \apj, 805,
  187, \dodoi{10.1088/0004-637X/805/2/187}

\bibitem[{{Chambers} {et~al.}(2016){Chambers}, {Magnier}, {Metcalfe},
  {Flewelling}, {Huber}, {Waters}, {Denneau}, {Draper}, {Farrow}, {Finkbeiner},
  {Holmberg}, {Koppenhoefer}, {Price}, {Rest}, {Saglia}, {Schlafly}, {Smartt},
  {Sweeney}, {Wainscoat}, {Burgett}, {Chastel}, {Grav}, {Heasley}, {Hodapp},
  {Jedicke}, {Kaiser}, {Kudritzki}, {Luppino}, {Lupton}, {Monet}, {Morgan},
  {Onaka}, {Shiao}, {Stubbs}, {Tonry}, {White}, {Ba{\~n}ados}, {Bell},
  {Bender}, {Bernard}, {Boegner}, {Boffi}, {Botticella}, {Calamida},
  {Casertano}, {Chen}, {Chen}, {Cole}, {Deacon}, {Frenk}, {Fitzsimmons},
  {Gezari}, {Gibbs}, {Goessl}, {Goggia}, {Gourgue}, {Goldman}, {Grant},
  {Grebel}, {Hambly}, {Hasinger}, {Heavens}, {Heckman}, {Henderson}, {Henning},
  {Holman}, {Hopp}, {Ip}, {Isani}, {Jackson}, {Keyes}, {Koekemoer}, {Kotak},
  {Le}, {Liska}, {Long}, {Lucey}, {Liu}, {Martin}, {Masci}, {McLean}, {Mindel},
  {Misra}, {Morganson}, {Murphy}, {Obaika}, {Narayan}, {Nieto-Santisteban},
  {Norberg}, {Peacock}, {Pier}, {Postman}, {Primak}, {Rae}, {Rai}, {Riess},
  {Riffeser}, {Rix}, {R{\"o}ser}, {Russel}, {Rutz}, {Schilbach}, {Schultz},
  {Scolnic}, {Strolger}, {Szalay}, {Seitz}, {Small}, {Smith}, {Soderblom},
  {Taylor}, {Thomson}, {Taylor}, {Thakar}, {Thiel}, {Thilker}, {Unger},
  {Urata}, {Valenti}, {Wagner}, {Walder}, {Walter}, {Watters}, {Werner},
  {Wood-Vasey}, \& {Wyse}}]{Chambers16}
{Chambers}, K.~C., {Magnier}, E.~A., {Metcalfe}, N., {et~al.} 2016, arXiv
  e-prints, arXiv:1612.05560.
\newblock \doarXiv{1612.05560}

\bibitem[{{Chandra} \& {Frail}(2012)}]{Chandra12}
{Chandra}, P., \& {Frail}, D.~A. 2012, \apj, 746, 156,
  \dodoi{10.1088/0004-637X/746/2/156}

\bibitem[{{Chen} \& {Shen}(2018)}]{Chen18}
{Chen}, J.-H., \& {Shen}, R.-F. 2018, \apj, 867, 20,
  \dodoi{10.3847/1538-4357/aadfda}

\bibitem[{{Chen} {et~al.}(2017){Chen}, {Smartt}, {Yates}, {Nicholl},
  {Kr{\"u}hler}, {Schady}, {Dennefeld}, \& {Inserra}}]{Chen17}
{Chen}, T.-W., {Smartt}, S.~J., {Yates}, R.~M., {et~al.} 2017, \mnras, 470,
  3566, \dodoi{10.1093/mnras/stx1428}

\bibitem[{{Chevalier}(1982)}]{Chevalier82b}
{Chevalier}, R.~A. 1982, \apj, 259, 302, \dodoi{10.1086/160167}

\bibitem[{{Chevalier}(1998)}]{Chevalier98}
---. 1998, \apj, 499, 810, \dodoi{10.1086/305676}

\bibitem[{{Chevalier} \& {Fransson}(2006)}]{Chevalier06}
{Chevalier}, R.~A., \& {Fransson}, C. 2006, \apj, 651, 381,
  \dodoi{10.1086/507606}

\bibitem[{{Chilingarian} {et~al.}(2015){Chilingarian}, {Beletsky}, {Moran},
  {Brown}, {McLeod}, \& {Fabricant}}]{2015PASP..127..406C}
{Chilingarian}, I., {Beletsky}, Y., {Moran}, S., {et~al.} 2015, \pasp, 127,
  406, \dodoi{10.1086/680598}

\bibitem[{{Chomiuk} {et~al.}(2012){Chomiuk}, {Soderberg}, {Margutti}, {Berger},
  {Milisavljevic}, \& {Sanders}}]{Chomiuk12b}
{Chomiuk}, L., {Soderberg}, A., {Margutti}, R., {et~al.} 2012, The Astronomer's
  Telegram, 3931, 1

\bibitem[{{Chonis} \& {Gaskell}(2008)}]{Chonis2008}
{Chonis}, T.~S., \& {Gaskell}, C.~M. 2008, \aj, 135, 264,
  \dodoi{10.1088/0004-6256/135/1/264}

\bibitem[{{Chornock} {et~al.}(2014){Chornock}, {Berger}, {Gezari}, {Zauderer},
  {Rest}, {Chomiuk}, {Kamble}, {Soderberg}, {Czekala}, {Dittmann}, {Drout},
  {Foley}, {Fong}, {Huber}, {Kirshner}, {Lawrence}, {Lunnan}, {Marion},
  {Narayan}, {Riess}, {Roth}, {Sanders}, {Scolnic}, {Smartt}, {Smith},
  {Stubbs}, {Tonry}, {Burgett}, {Chambers}, {Flewelling}, {Hodapp}, {Kaiser},
  {Magnier}, {Martin}, {Neill}, {Price}, \& {Wainscoat}}]{Chornock14}
{Chornock}, R., {Berger}, E., {Gezari}, S., {et~al.} 2014, \apj, 780, 44,
  \dodoi{10.1088/0004-637X/780/1/44}

\bibitem[{{Condon} {et~al.}(1998){Condon}, {Cotton}, {Greisen}, {Yin},
  {Perley}, {Taylor}, \& {Broderick}}]{Condon98}
{Condon}, J.~J., {Cotton}, W.~D., {Greisen}, E.~W., {et~al.} 1998, \aj, 115,
  1693, \dodoi{10.1086/300337}

\bibitem[{{Conroy} \& {Gunn}(2010)}]{Conroy10}
{Conroy}, C., \& {Gunn}, J.~E. 2010, \apj, 712, 833,
  \dodoi{10.1088/0004-637X/712/2/833}

\bibitem[{{Coppejans} {et~al.}(2018){Coppejans}, {Margutti}, {Guidorzi},
  {Chomiuk}, {Alexander}, {Berger}, {Bietenholz}, {Blanchard}, {Challis},
  {Chornock}, {Drout}, {Fong}, {MacFadyen}, {Migliori}, {Milisavljevic},
  {Nicholl}, {Parrent}, {Terreran}, \& {Zauderer}}]{Coppejans18}
{Coppejans}, D.~L., {Margutti}, R., {Guidorzi}, C., {et~al.} 2018, \apj, 856,
  56, \dodoi{10.3847/1538-4357/aab36e}

\bibitem[{{Corsi} {et~al.}(2017){Corsi}, {Cenko}, {Kasliwal}, {Quimby},
  {Kulkarni}, {Frail}, {Goldstein}, {Blagorodnova}, {Connaughton}, {Perley},
  {Singer}, {Copperwheat}, {Fremling}, {Kupfer}, {Piascik}, {Steele}, {Taddia},
  {Vedantham}, {Kutyrev}, {Palliyaguru}, {Roberts}, {Sollerman}, {Troja}, \&
  {Veilleux}}]{Corsi17}
{Corsi}, A., {Cenko}, S.~B., {Kasliwal}, M.~M., {et~al.} 2017, \apj, 847, 54,
  \dodoi{10.3847/1538-4357/aa85e5}

\bibitem[{{Coughlin}(2019)}]{Coughlin19}
{Coughlin}, E.~R. 2019, \apj, 880, 108, \dodoi{10.3847/1538-4357/ab29e6}

\bibitem[{{Darbha} {et~al.}(2019){Darbha}, {Coughlin}, {Kasen}, \&
  {Nixon}}]{Darbha19}
{Darbha}, S., {Coughlin}, E.~R., {Kasen}, D., \& {Nixon}, C. 2019, \mnras, 488,
  5267, \dodoi{10.1093/mnras/stz1923}

\bibitem[{{Dermer} {et~al.}(2000){Dermer}, {Chiang}, \& {Mitman}}]{Dermer2000}
{Dermer}, C.~D., {Chiang}, J., \& {Mitman}, K.~E. 2000, \apj, 537, 785,
  \dodoi{10.1086/309061}

\bibitem[{{Dey} {et~al.}(2019){Dey}, {Schlegel}, {Lang}, {Blum}, {Burleigh},
  {Fan}, {Findlay}, {Finkbeiner}, {Herrera}, {Juneau}, {Landriau}, {Levi},
  {McGreer}, {Meisner}, {Myers}, {Moustakas}, {Nugent}, {Patej}, {Schlafly},
  {Walker}, {Valdes}, {Weaver}, {Y{\`e}che}, {Zou}, {Zhou}, {Abareshi},
  {Abbott}, {Abolfathi}, {Aguilera}, {Alam}, {Allen}, {Alvarez}, {Annis},
  {Ansarinejad}, {Aubert}, {Beechert}, {Bell}, {BenZvi}, {Beutler}, {Bielby},
  {Bolton}, {Brice{\~n}o}, {Buckley-Geer}, {Butler}, {Calamida}, {Carlberg},
  {Carter}, {Casas}, {Castander}, {Choi}, {Comparat}, {Cukanovaite}, {Delubac},
  {DeVries}, {Dey}, {Dhungana}, {Dickinson}, {Ding}, {Donaldson}, {Duan},
  {Duckworth}, {Eftekharzadeh}, {Eisenstein}, {Etourneau}, {Fagrelius},
  {Farihi}, {Fitzpatrick}, {Font-Ribera}, {Fulmer}, {G{\"a}nsicke},
  {Gaztanaga}, {George}, {Gerdes}, {Gontcho}, {Gorgoni}, {Green}, {Guy},
  {Harmer}, {Hernand ez}, {Honscheid}, {Huang}, {James}, {Jannuzi}, {Jiang},
  {Joyce}, {Karcher}, {Karkar}, {Kehoe}, {Kneib}, {Kueter-Young}, {Lan},
  {Lauer}, {Le Guillou}, {Le Van Suu}, {Lee}, {Lesser}, {Perreault Levasseur},
  {Li}, {Mann}, {Marshall}, {Mart{\'\i}nez-V{\'a}zquez}, {Martini}, {du Mas des
  Bourboux}, {McManus}, {Meier}, {M{\'e}nard}, {Metcalfe},
  {Mu{\~n}oz-Guti{\'e}rrez}, {Najita}, {Napier}, {Narayan}, {Newman}, {Nie},
  {Nord}, {Norman}, {Olsen}, {Paat}, {Palanque-Delabrouille}, {Peng},
  {Poppett}, {Poremba}, {Prakash}, {Rabinowitz}, {Raichoor}, {Rezaie},
  {Robertson}, {Roe}, {Ross}, {Ross}, {Rudnick}, {Safonova}, {Saha},
  {S{\'a}nchez}, {Savary}, {Schweiker}, {Scott}, {Seo}, {Shan}, {Silva},
  {Slepian}, {Soto}, {Sprayberry}, {Staten}, {Stillman}, {Stupak}, {Summers},
  {Sien Tie}, {Tirado}, {Vargas-Maga{\~n}a}, {Vivas}, {Wechsler}, {Williams},
  {Yang}, {Yang}, {Yapici}, {Zaritsky}, {Zenteno}, {Zhang}, {Zhang}, {Zhou}, \&
  {Zhou}}]{Dey19}
{Dey}, A., {Schlegel}, D.~J., {Lang}, D., {et~al.} 2019, \aj, 157, 168,
  \dodoi{10.3847/1538-3881/ab089d}

\bibitem[{{Drake} {et~al.}(2009){Drake}, {Djorgovski}, {Mahabal}, {Beshore},
  {Larson}, {Graham}, {Williams}, {Christensen}, {Catelan}, {Boattini},
  {Gibbs}, {Hill}, \& {Kowalski}}]{Drake09}
{Drake}, A.~J., {Djorgovski}, S.~G., {Mahabal}, A., {et~al.} 2009, \apj, 696,
  870, \dodoi{10.1088/0004-637X/696/1/870}

\bibitem[{{Drout} {et~al.}(2013){Drout}, {Soderberg}, {Mazzali}, {Parrent},
  {Margutti}, {Milisavljevic}, {Sand ers}, {Chornock}, {Foley}, {Kirshner},
  {Filippenko}, {Li}, {Brown}, {Cenko}, {Chakraborti}, {Challis}, {Friedman},
  {Ganeshalingam}, {Hicken}, {Jensen}, {Modjaz}, {Perets}, {Silverman}, \&
  {Wong}}]{Drout13}
{Drout}, M.~R., {Soderberg}, A.~M., {Mazzali}, P.~A., {et~al.} 2013, \apj, 774,
  58, \dodoi{10.1088/0004-637X/774/1/58}

\bibitem[{{Drout} {et~al.}(2014){Drout}, {Chornock}, {Soderberg}, {Sand ers},
  {McKinnon}, {Rest}, {Foley}, {Milisavljevic}, {Margutti}, {Berger},
  {Calkins}, {Fong}, {Gezari}, {Huber}, {Kankare}, {Kirshner}, {Leibler},
  {Lunnan}, {Mattila}, {Marion}, {Narayan}, {Riess}, {Roth}, {Scolnic},
  {Smartt}, {Tonry}, {Burgett}, {Chambers}, {Hodapp}, {Jedicke}, {Kaiser},
  {Magnier}, {Metcalfe}, {Morgan}, {Price}, \& {Waters}}]{Drout14}
{Drout}, M.~R., {Chornock}, R., {Soderberg}, A.~M., {et~al.} 2014, \apj, 794,
  23, \dodoi{10.1088/0004-637X/794/1/23}

\bibitem[{{Eftekhari} {et~al.}(2018){Eftekhari}, {Berger}, {Zauderer},
  {Margutti}, \& {Alexander}}]{Eftekhari18}
{Eftekhari}, T., {Berger}, E., {Zauderer}, B.~A., {Margutti}, R., \&
  {Alexander}, K.~D. 2018, \apj, 854, 86, \dodoi{10.3847/1538-4357/aaa8e0}

\bibitem[{{Eftekhari} {et~al.}(2019){Eftekhari}, {Berger}, {Margalit},
  {Blanchard}, {Patton}, {Demorest}, {Williams}, {Chatterjee}, {Cordes},
  {Lunnan}, {Metzger}, \& {Nicholl}}]{Eftekhari19}
{Eftekhari}, T., {Berger}, E., {Margalit}, B., {et~al.} 2019, \apjl, 876, L10,
  \dodoi{10.3847/2041-8213/ab18a5}

\bibitem[{{Frohmaier} {et~al.}(2018){Frohmaier}, {Sullivan}, {Maguire}, \&
  {Nugent}}]{Frohmaier18}
{Frohmaier}, C., {Sullivan}, M., {Maguire}, K., \& {Nugent}, P. 2018, \apj,
  858, 50, \dodoi{10.3847/1538-4357/aabc0b}

\bibitem[{{Frohmaier} {et~al.}(2017){Frohmaier}, {Sullivan}, {Nugent},
  {Goldstein}, \& {DeRose}}]{Frohmaier17}
{Frohmaier}, C., {Sullivan}, M., {Nugent}, P.~E., {Goldstein}, D.~A., \&
  {DeRose}, J. 2017, \apjs, 230, 4, \dodoi{10.3847/1538-4365/aa6d70}

\bibitem[{{Granot} \& {Sari}(2002)}]{Granot02}
{Granot}, J., \& {Sari}, R. 2002, \apj, 568, 820, \dodoi{10.1086/338966}

\bibitem[{{Hinkle} {et~al.}(2020){Hinkle}, {Holoien}, {Shappee}, {Auchettl},
  {Kochanek}, {Stanek}, {Payne}, \& {Thompson}}]{Hinkle20}
{Hinkle}, J.~T., {Holoien}, T. W.~S., {Shappee}, B.~J., {et~al.} 2020, arXiv
  e-prints, arXiv:2001.08215.
\newblock \doarXiv{2001.08215}

\bibitem[{{Ho} {et~al.}(2019){Ho}, {Phinney}, {Ravi}, {Kulkarni}, {Petitpas},
  {Emonts}, {Bhalerao}, {Blundell}, {Cenko}, {Dobie}, {Howie}, {Kamraj},
  {Kasliwal}, {Murphy}, {Perley}, {Sridharan}, \& {Yoon}}]{Ho19}
{Ho}, A. Y.~Q., {Phinney}, E.~S., {Ravi}, V., {et~al.} 2019, \apj, 871, 73,
  \dodoi{10.3847/1538-4357/aaf473}

\bibitem[{{Ho} {et~al.}(2020){Ho}, {Perley}, {Kulkarni}, {Andreoni}, {Bellm},
  {Burdge}, {Chand ra}, {Coughlin}, {De}, {Dong}, {Golkhou}, {Graham},
  {Fredericks}, {Helou}, {Horesh}, {Laher}, {Masci}, {Ridnaia}, {Rusholme},
  {Shupe}, \& {Svinkin}}]{Ho20}
{Ho}, A. Y.~Q., {Perley}, D.~A., {Kulkarni}, S.~R., {et~al.} 2020, arXiv
  e-prints, arXiv:2003.01222.
\newblock \doarXiv{2003.01222}

\bibitem[{{Holoien} {et~al.}(2016){Holoien}, {Kochanek}, {Prieto}, {Stanek},
  {Dong}, {Shappee}, {Grupe}, {Brown}, {Basu}, {Beacom}, {Bersier},
  {Brimacombe}, {Danilet}, {Falco}, {Guo}, {Jose}, {Herczeg}, {Long},
  {Pojmanski}, {Simonian}, {Szczygie{\l}}, {Thompson}, {Thorstensen}, {Wagner},
  \& {Wo{\'z}niak}}]{Holoien16}
{Holoien}, T.~W.~S., {Kochanek}, C.~S., {Prieto}, J.~L., {et~al.} 2016, \mnras,
  455, 2918, \dodoi{10.1093/mnras/stv2486}

\bibitem[{{Hotokezaka} {et~al.}(2017){Hotokezaka}, {Kashiyama}, \&
  {Murase}}]{Hotokezaka17}
{Hotokezaka}, K., {Kashiyama}, K., \& {Murase}, K. 2017, \apj, 850, 18,
  \dodoi{10.3847/1538-4357/aa8c7d}

\bibitem[{{Huang} {et~al.}(2002){Huang}, {Dai}, \& {Lu}}]{Huang02}
{Huang}, Y.~F., {Dai}, Z.~G., \& {Lu}, T. 2002, \mnras, 332, 735,
  \dodoi{10.1046/j.1365-8711.2002.05334.x}

\bibitem[{{Jarrett} {et~al.}(2017){Jarrett}, {Cluver}, {Magoulas}, {Bilicki},
  {Alpaslan}, {Bland-Hawthorn}, {Brough}, {Brown}, {Croom}, {Driver},
  {Holwerda}, {Hopkins}, {Loveday}, {Norberg}, {Peacock}, {Popescu}, {Sadler},
  {Taylor}, {Tuffs}, \& {Wang}}]{Jarrett17}
{Jarrett}, T.~H., {Cluver}, M.~E., {Magoulas}, C., {et~al.} 2017, \apj, 836,
  182, \dodoi{10.3847/1538-4357/836/2/182}

\bibitem[{{Kalberla} {et~al.}(2005){Kalberla}, {Burton}, {Hartmann}, {Arnal},
  {Bajaja}, {Morras}, \& {P{\"o}ppel}}]{Kalberla05}
{Kalberla}, P.~M.~W., {Burton}, W.~B., {Hartmann}, D., {et~al.} 2005, \aap,
  440, 775, \dodoi{10.1051/0004-6361:20041864}

\bibitem[{{Kauffmann} {et~al.}(2003){Kauffmann}, {Heckman}, {Tremonti},
  {Brinchmann}, {Charlot}, {White}, {Ridgway}, {Brinkmann}, {Fukugita}, \&
  {Hall}}]{Kauffmann03}
{Kauffmann}, G., {Heckman}, T.~M., {Tremonti}, C., {et~al.} 2003, \mnras, 346,
  1055, \dodoi{10.1111/j.1365-2966.2003.07154.x}

\bibitem[{{Kewley} \& {Dopita}(2002)}]{Kewley02}
{Kewley}, L.~J., \& {Dopita}, M.~A. 2002, \apjs, 142, 35,
  \dodoi{10.1086/341326}

\bibitem[{{Kriek} {et~al.}(2009){Kriek}, {van Dokkum}, {Labb{\'e}}, {Franx},
  {Illingworth}, {Marchesini}, \& {Quadri}}]{Kriek09}
{Kriek}, M., {van Dokkum}, P.~G., {Labb{\'e}}, I., {et~al.} 2009, \apj, 700,
  221, \dodoi{10.1088/0004-637X/700/1/221}

\bibitem[{{Krolik} {et~al.}(2016){Krolik}, {Piran}, {Svirski}, \&
  {Cheng}}]{Krolik16}
{Krolik}, J., {Piran}, T., {Svirski}, G., \& {Cheng}, R.~M. 2016, \apj, 827,
  127, \dodoi{10.3847/0004-637X/827/2/127}

\bibitem[{{Kuin} {et~al.}(2019){Kuin}, {Wu}, {Oates}, {Lien}, {Emery},
  {Kennea}, {de Pasquale}, {Han}, {Brown}, \& {Tohuvavohu}}]{Kuin19}
{Kuin}, N. P.~M., {Wu}, K., {Oates}, S., {et~al.} 2019, \mnras, 487, 2505,
  \dodoi{10.1093/mnras/stz053}

\bibitem[{{Lang} {et~al.}(2016){Lang}, {Hogg}, \& {Mykytyn}}]{Lang16}
{Lang}, D., {Hogg}, D.~W., \& {Mykytyn}, D. 2016, {The Tractor: Probabilistic
  astronomical source detection and measurement}.
\newblock \doeprint{1604.008}

\bibitem[{{Law} {et~al.}(2019){Law}, {Omand}, {Kashiyama}, {Murase}, {Bower},
  {Aggarwal}, {Burke-Spolaor}, {Butler}, {Demorest}, {Lazio}, {Linford},
  {Tendulkar}, \& {Rupen}}]{Law19}
{Law}, C.~J., {Omand}, C.~M.~B., {Kashiyama}, K., {et~al.} 2019, \apj, 886, 24,
  \dodoi{10.3847/1538-4357/ab4adb}

\bibitem[{{Law} {et~al.}(2009){Law}, {Kulkarni}, {Ofek}, {Quimby}, {Kasliwal},
  \& {Palomar Transient Factory Collaboration}}]{Law09PTF}
{Law}, N.~M., {Kulkarni}, S., {Ofek}, E., {et~al.} 2009, in Bulletin of the
  American Astronomical Society, Vol.~41, American Astronomical Society Meeting
  Abstracts 213, 469.01

\bibitem[{{Law-Smith} {et~al.}(2017){Law-Smith}, {Ramirez-Ruiz}, {Ellison}, \&
  {Foley}}]{Law-Smith2017}
{Law-Smith}, J., {Ramirez-Ruiz}, E., {Ellison}, S.~L., \& {Foley}, R.~J. 2017,
  \apj, 850, 22, \dodoi{10.3847/1538-4357/aa94c7}

\bibitem[{{Lazzati} {et~al.}(2012){Lazzati}, {Morsony}, {Blackwell}, \&
  {Begelman}}]{Lazzati12}
{Lazzati}, D., {Morsony}, B.~J., {Blackwell}, C.~H., \& {Begelman}, M.~C. 2012,
  \apj, 750, 68, \dodoi{10.1088/0004-637X/750/1/68}

\bibitem[{{Li} {et~al.}(2011{\natexlab{a}}){Li}, {Chornock}, {Leaman},
  {Filippenko}, {Poznanski}, {Wang}, {Ganeshalingam}, \&
  {Mannucci}}]{Li11_LOSS}
{Li}, W., {Chornock}, R., {Leaman}, J., {et~al.} 2011{\natexlab{a}}, \mnras,
  412, 1473, \dodoi{10.1111/j.1365-2966.2011.18162.x}

\bibitem[{{Li} {et~al.}(2011{\natexlab{b}}){Li}, {Bloom}, {Podsiadlowski},
  {Miller}, {Cenko}, {Jha}, {Sullivan}, {Howell}, {Nugent}, {Butler}, {Ofek},
  {Kasliwal}, {Richards}, {Stockton}, {Shih}, {Bildsten}, {Shara}, {Bibby},
  {Filippenko}, {Ganeshalingam}, {Silverman}, {Kulkarni}, {Law}, {Poznanski},
  {Quimby}, {McCully}, {Patel}, {Maguire}, \& {Shen}}]{Li11}
{Li}, W., {Bloom}, J.~S., {Podsiadlowski}, P., {et~al.} 2011{\natexlab{b}},
  \nat, 480, 348, \dodoi{10.1038/nature10646}

\bibitem[{{Lunnan} {et~al.}(2014){Lunnan}, {Chornock}, {Berger}, {Laskar},
  {Fong}, {Rest}, {Sanders}, {Challis}, {Drout}, {Foley}, {Huber}, {Kirshner},
  {Leibler}, {Marion}, {McCrum}, {Milisavljevic}, {Narayan}, {Scolnic},
  {Smartt}, {Smith}, {Soderberg}, {Tonry}, {Burgett}, {Chambers}, {Flewelling},
  {Hodapp}, {Kaiser}, {Magnier}, {Price}, \& {Wainscoat}}]{Lunnan14}
{Lunnan}, R., {Chornock}, R., {Berger}, E., {et~al.} 2014, \apj, 787, 138,
  \dodoi{10.1088/0004-637X/787/2/138}

\bibitem[{{Lyutikov} \& {Toonen}(2019)}]{Lyutikov19}
{Lyutikov}, M., \& {Toonen}, S. 2019, \mnras, 487, 5618,
  \dodoi{10.1093/mnras/stz1640}

\bibitem[{{MacFadyen} \& {Woosley}(1999)}]{MacFadyen99}
{MacFadyen}, A.~I., \& {Woosley}, S.~E. 1999, \apj, 524, 262,
  \dodoi{10.1086/307790}

\bibitem[{{MacFadyen} {et~al.}(2001){MacFadyen}, {Woosley}, \&
  {Heger}}]{MacFadyen01}
{MacFadyen}, A.~I., {Woosley}, S.~E., \& {Heger}, A. 2001, \apj, 550, 410,
  \dodoi{10.1086/319698}

\bibitem[{{Margutti} {et~al.}(2013{\natexlab{a}}){Margutti}, {Soderberg},
  {Wieringa}, {Edwards}, {Chevalier}, {Morsony}, {Barniol Duran}, {Sironi},
  {Zauderer}, {Milisavljevic}, {Kamble}, \& {Pian}}]{Margutti13b}
{Margutti}, R., {Soderberg}, A.~M., {Wieringa}, M.~H., {et~al.}
  2013{\natexlab{a}}, \apj, 778, 18, \dodoi{10.1088/0004-637X/778/1/18}

\bibitem[{{Margutti} {et~al.}(2013{\natexlab{b}}){Margutti}, {Zaninoni},
  {Bernardini}, {Chincarini}, {Pasotti}, {Guidorzi}, {Angelini}, {Burrows},
  {Capalbi}, {Evans}, {Gehrels}, {Kennea}, {Mangano}, {Moretti}, {Nousek},
  {Osborne}, {Page}, {Perri}, {Racusin}, {Romano}, {Sbarufatti}, {Stafford}, \&
  {Stamatikos}}]{Margutti13}
{Margutti}, R., {Zaninoni}, E., {Bernardini}, M.~G., {et~al.}
  2013{\natexlab{b}}, \mnras, 428, 729, \dodoi{10.1093/mnras/sts066}

\bibitem[{{Margutti} {et~al.}(2014{\natexlab{a}}){Margutti}, {Milisavljevic},
  {Soderberg}, {Chornock}, {Zauderer}, {Murase}, {Guidorzi}, {Sanders}, {Kuin},
  {Fransson}, {Levesque}, {Chandra}, {Berger}, {Bianco}, {Brown}, {Challis},
  {Chatzopoulos}, {Cheung}, {Choi}, {Chomiuk}, {Chugai}, {Contreras}, {Drout},
  {Fesen}, {Foley}, {Fong}, {Friedman}, {Gall}, {Gehrels}, {Hjorth}, {Hsiao},
  {Kirshner}, {Im}, {Leloudas}, {Lunnan}, {Marion}, {Martin}, {Morrell},
  {Neugent}, {Omodei}, {Phillips}, {Rest}, {Silverman}, {Strader},
  {Stritzinger}, {Szalai}, {Utterback}, {Vinko}, {Wheeler}, {Arnett},
  {Campana}, {Chevalier}, {Ginsburg}, {Kamble}, {Roming}, {Pritchard}, \&
  {Stringfellow}}]{Margutti14}
{Margutti}, R., {Milisavljevic}, D., {Soderberg}, A.~M., {et~al.}
  2014{\natexlab{a}}, \apj, 780, 21, \dodoi{10.1088/0004-637X/780/1/21}

\bibitem[{{Margutti} {et~al.}(2014{\natexlab{b}}){Margutti}, {Milisavljevic},
  {Soderberg}, {Guidorzi}, {Morsony}, {Sanders}, {Chakraborti}, {Ray},
  {Kamble}, {Drout}, {Parrent}, {Zauderer}, \& {Chomiuk}}]{Margutti14b}
---. 2014{\natexlab{b}}, \apj, 797, 107, \dodoi{10.1088/0004-637X/797/2/107}

\bibitem[{{Margutti} {et~al.}(2017{\natexlab{a}}){Margutti}, {Metzger},
  {Chornock}, {Milisavljevic}, {Berger}, {Blanchard}, {Guidorzi}, {Migliori},
  {Kamble}, {Lunnan}, {Nicholl}, {Coppejans}, {Dall'Osso}, {Drout}, {Perna}, \&
  {Sbarufatti}}]{Margutti17b}
{Margutti}, R., {Metzger}, B.~D., {Chornock}, R., {et~al.} 2017{\natexlab{a}},
  \apj, 836, 25, \dodoi{10.3847/1538-4357/836/1/25}

\bibitem[{{Margutti} {et~al.}(2017{\natexlab{b}}){Margutti}, {Kamble},
  {Milisavljevic}, {Zapartas}, {de Mink}, {Drout}, {Chornock}, {Risaliti},
  {Zauderer}, {Bietenholz}, {Cantiello}, {Chakraborti}, {Chomiuk}, {Fong},
  {Grefenstette}, {Guidorzi}, {Kirshner}, {Parrent}, {Patnaude}, {Soderberg},
  {Gehrels}, \& {Harrison}}]{Margutti17}
{Margutti}, R., {Kamble}, A., {Milisavljevic}, D., {et~al.} 2017{\natexlab{b}},
  \apj, 835, 140, \dodoi{10.3847/1538-4357/835/2/140}

\bibitem[{{Margutti} {et~al.}(2019){Margutti}, {Metzger}, {Chornock}, {Vurm},
  {Roth}, {Grefenstette}, {Savchenko}, {Cartier}, {Steiner}, {Terreran},
  {Margalit}, {Migliori}, {Milisavljevic}, {Alexand er}, {Bietenholz},
  {Blanchard}, {Bozzo}, {Brethauer}, {Chilingarian}, {Coppejans}, {Ducci},
  {Ferrigno}, {Fong}, {G{\"o}tz}, {Guidorzi}, {Hajela}, {Hurley}, {Kuulkers},
  {Laurent}, {Mereghetti}, {Nicholl}, {Patnaude}, {Ubertini}, {Banovetz},
  {Bartel}, {Berger}, {Coughlin}, {Eftekhari}, {Frederiks}, {Kozlova},
  {Laskar}, {Svinkin}, {Drout}, {MacFadyen}, \& {Paterson}}]{Margutti19}
{Margutti}, R., {Metzger}, B.~D., {Chornock}, R., {et~al.} 2019, \apj, 872, 18,
  \dodoi{10.3847/1538-4357/aafa01}

\bibitem[{{Marleau} {et~al.}(2013){Marleau}, {Clancy}, \&
  {Bianconi}}]{Marleau13}
{Marleau}, F.~R., {Clancy}, D., \& {Bianconi}, M. 2013, \mnras, 435, 3085,
  \dodoi{10.1093/mnras/stt1503}

\bibitem[{{Matheson} {et~al.}(2000){Matheson}, {Filippenko}, {Chornock},
  {Leonard}, \& {Li}}]{Matheson00}
{Matheson}, T., {Filippenko}, A.~V., {Chornock}, R., {Leonard}, D.~C., \& {Li},
  W. 2000, \aj, 119, 2303, \dodoi{10.1086/301352}

\bibitem[{{Mattila} {et~al.}(2018){Mattila}, {P{\'e}rez-Torres}, {Efstathiou},
  {Mimica}, {Fraser}, {Kankare}, {Alberdi}, {Aloy}, {Heikkil{\"a}}, {Jonker},
  {Lundqvist}, {Mart{\'\i}-Vidal}, {Meikle}, {Romero-Ca{\~n}izales}, {Smartt},
  {Tsygankov}, {Varenius}, {Alonso-Herrero}, {Bondi}, {Fransson},
  {Herrero-Illana}, {Kangas}, {Kotak}, {Ram{\'\i}rez-Olivencia},
  {V{\"a}is{\"a}nen}, {Beswick}, {Clements}, {Greimel}, {Harmanen},
  {Kotilainen}, {Nandra}, {Reynolds}, {Ryder}, {Walton}, {Wiik}, \&
  {{\"O}stlin}}]{Mattila18}
{Mattila}, S., {P{\'e}rez-Torres}, M., {Efstathiou}, A., {et~al.} 2018,
  Science, 361, 482, \dodoi{10.1126/science.aao4669}

\bibitem[{{McLeod} {et~al.}(2012){McLeod}, {Fabricant}, {Nystrom}, {McCracken},
  {Amato}, {Bergner}, {Brown}, {Burke}, {Chilingarian}, {Conroy}, {Curley},
  {Furesz}, {Geary}, {Hertz}, {Holwell}, {Matthews}, {Norton}, {Park}, {Roll},
  {Zajac}, {Epps}, \& {Martini}}]{2012PASP..124.1318M}
{McLeod}, B., {Fabricant}, D., {Nystrom}, G., {et~al.} 2012, \pasp, 124, 1318,
  \dodoi{10.1086/669044}

\bibitem[{{McMullin} {et~al.}(2007){McMullin}, {Waters}, {Schiebel}, {Young},
  \& {Golap}}]{McMullin07}
{McMullin}, J.~P., {Waters}, B., {Schiebel}, D., {Young}, W., \& {Golap}, K.
  2007, in Astronomical Society of the Pacific Conference Series, Vol. 376,
  Astronomical Data Analysis Software and Systems XVI, ed. R.~A. {Shaw},
  F.~{Hill}, \& D.~J. {Bell}, 127

\bibitem[{{Meisner} {et~al.}(2018){Meisner}, {Lang}, \&
  {Schlegel}}]{meisner2018}
{Meisner}, A.~M., {Lang}, D., \& {Schlegel}, D.~J. 2018, Research Notes of the
  American Astronomical Society, 2, 1, \dodoi{10.3847/2515-5172/aaa4bc}

\bibitem[{{Mezcua} {et~al.}(2018){Mezcua}, {Civano}, {Marchesi}, {Suh},
  {Fabbiano}, \& {Volonteri}}]{Mezcua18}
{Mezcua}, M., {Civano}, F., {Marchesi}, S., {et~al.} 2018, \mnras, 478, 2576,
  \dodoi{10.1093/mnras/sty1163}

\bibitem[{{Micha{\l}owski} {et~al.}(2018){Micha{\l}owski}, {Gentile},
  {Kr{\"u}hler}, {Kuncarayakti}, {Kamphuis}, {Hjorth}, {Berta}, {D'Elia},
  {Elliott}, {Galbany}, {Greiner}, {Hunt}, {Koprowski}, {Le Floc'h}, {Nicuesa
  Guelbenzu}, {Palazzi}, {Rasmussen}, {Rossi}, {Savaglio}, {de Ugarte Postigo},
  {van der Werf}, \& {Vergani}}]{Michalowski18}
{Micha{\l}owski}, M.~J., {Gentile}, G., {Kr{\"u}hler}, T., {et~al.} 2018, \aap,
  618, A104, \dodoi{10.1051/0004-6361/201732356}

\bibitem[{{Milisavljevic} {et~al.}(2015){Milisavljevic}, {Margutti}, {Kamble},
  {Patnaude}, {Raymond}, {Eldridge}, {Fong}, {Bietenholz}, {Challis},
  {Chornock}, {Drout}, {Fransson}, {Fesen}, {Grindlay}, {Kirshner}, {Lunnan},
  {Mackey}, {Miller}, {Parrent}, {Sand ers}, {Soderberg}, \&
  {Zauderer}}]{Milisavljevic15}
{Milisavljevic}, D., {Margutti}, R., {Kamble}, A., {et~al.} 2015, \apj, 815,
  120, \dodoi{10.1088/0004-637X/815/2/120}

\bibitem[{{Mohan} \& {Rafferty}(2015)}]{Mohan15}
{Mohan}, N., \& {Rafferty}, D. 2015, {PyBDSF: Python Blob Detection and Source
  Finder}, Astrophysics Source Code Library.
\newblock \doeprint{1502.007}

\bibitem[{{Moriya} {et~al.}(2017){Moriya}, {Mazzali}, {Tominaga}, {Hachinger},
  {Blinnikov}, {Tauris}, {Takahashi}, {Tanaka}, {Langer}, \&
  {Podsiadlowski}}]{Moriya17}
{Moriya}, T.~J., {Mazzali}, P.~A., {Tominaga}, N., {et~al.} 2017, \mnras, 466,
  2085, \dodoi{10.1093/mnras/stw3225}

\bibitem[{{Nakar} \& {Piran}(2003)}]{Nakar03}
{Nakar}, E., \& {Piran}, T. 2003, \na, 8, 141,
  \dodoi{10.1016/S1384-1076(02)00202-6}

\bibitem[{{Nakar} \& {Sari}(2012)}]{Nakar12}
{Nakar}, E., \& {Sari}, R. 2012, \apj, 747, 88,
  \dodoi{10.1088/0004-637X/747/2/88}

\bibitem[{{Nava} {et~al.}(2012){Nava}, {Salvaterra}, {Ghirlanda}, {Ghisellini},
  {Campana}, {Covino}, {Cusumano}, {D'Avanzo}, {D'Elia}, {Fugazza}, {Melandri},
  {Sbarufatti}, {Vergani}, \& {Tagliaferri}}]{Nava12}
{Nava}, L., {Salvaterra}, R., {Ghirlanda}, G., {et~al.} 2012, \mnras, 421,
  1256, \dodoi{10.1111/j.1365-2966.2011.20394.x}

\bibitem[{{Nicholl} {et~al.}(2016){Nicholl}, {Berger}, {Smartt}, {Margutti},
  {Kamble}, {Alexander}, {Chen}, {Inserra}, {Arcavi}, {Blanchard}, {Cartier},
  {Chambers}, {Childress}, {Chornock}, {Cowperthwaite}, {Drout}, {Flewelling},
  {Fraser}, {Gal-Yam}, {Galbany}, {Harmanen}, {Holoien}, {Hosseinzadeh},
  {Howell}, {Huber}, {Jerkstrand }, {Kankare}, {Kochanek}, {Lin}, {Lunnan},
  {Magnier}, {Maguire}, {McCully}, {McDonald}, {Metzger}, {Milisavljevic},
  {Mitra}, {Reynolds}, {Saario}, {Shappee}, {Smith}, {Valenti}, {Villar},
  {Waters}, \& {Young}}]{Nicholl16}
{Nicholl}, M., {Berger}, E., {Smartt}, S.~J., {et~al.} 2016, \apj, 826, 39,
  \dodoi{10.3847/0004-637X/826/1/39}

\bibitem[{{Ofek} {et~al.}(2010){Ofek}, {Rabinak}, {Neill}, {Arcavi}, {Cenko},
  {Waxman}, {Kulkarni}, {Gal-Yam}, {Nugent}, {Bildsten}, {Bloom}, {Filippenko},
  {Forster}, {Howell}, {Jacobsen}, {Kasliwal}, {Law}, {Martin}, {Poznanski},
  {Quimby}, {Shen}, {Sullivan}, {Dekany}, {Rahmer}, {Hale}, {Smith},
  {Zolkower}, {Velur}, {Walters}, {Henning}, {Bui}, \& {McKenna}}]{Ofek10}
{Ofek}, E.~O., {Rabinak}, I., {Neill}, J.~D., {et~al.} 2010, \apj, 724, 1396,
  \dodoi{10.1088/0004-637X/724/2/1396}

\bibitem[{{Pasham} \& {van Velzen}(2018)}]{Pasham18}
{Pasham}, D.~R., \& {van Velzen}, S. 2018, \apj, 856, 1,
  \dodoi{10.3847/1538-4357/aab361}

\bibitem[{{Perets} {et~al.}(2010){Perets}, {Gal-Yam}, {Mazzali}, {Arnett},
  {Kagan}, {Filippenko}, {Li}, {Arcavi}, {Cenko}, {Fox}, {Leonard}, {Moon},
  {Sand}, {Soderberg}, {Anderson}, {James}, {Foley}, {Ganeshalingam}, {Ofek},
  {Bildsten}, {Nelemans}, {Shen}, {Weinberg}, {Metzger}, {Piro}, {Quataert},
  {Kiewe}, \& {Poznanski}}]{Perets10}
{Perets}, H.~B., {Gal-Yam}, A., {Mazzali}, P.~A., {et~al.} 2010, \nat, 465,
  322, \dodoi{10.1038/nature09056}

\bibitem[{{Perley} {et~al.}(2019){Perley}, {Mazzali}, {Yan}, {Cenko}, {Gezari},
  {Taggart}, {Blagorodnova}, {Fremling}, {Mockler}, {Singh}, {Tominaga},
  {Tanaka}, {Watson}, {Ahumada}, {Anupama}, {Ashall}, {Becerra}, {Bersier},
  {Bhalerao}, {Bloom}, {Butler}, {Copperwheat}, {Coughlin}, {De}, {Drake},
  {Duev}, {Frederick}, {Gonz{\'a}lez}, {Goobar}, {Heida}, {Ho}, {Horst},
  {Hung}, {Itoh}, {Jencson}, {Kasliwal}, {Kawai}, {Khanam}, {Kulkarni},
  {Kumar}, {Kumar}, {Kutyrev}, {Lee}, {Maeda}, {Mahabal}, {Murata}, {Neill},
  {Ngeow}, {Penprase}, {Pian}, {Quimby}, {Ramirez-Ruiz}, {Richer},
  {Rom{\'a}n-Z{\'u}{\~n}iga}, {Sahu}, {Srivastav}, {Socia}, {Sollerman},
  {Tachibana}, {Taddia}, {Tinyanont}, {Troja}, {Ward}, {Wee}, \&
  {Yu}}]{Perley19}
{Perley}, D.~A., {Mazzali}, P.~A., {Yan}, L., {et~al.} 2019, \mnras, 484, 1031,
  \dodoi{10.1093/mnras/sty3420}

\bibitem[{{Poznanski} {et~al.}(2010){Poznanski}, {Chornock}, {Nugent}, {Bloom},
  {Filippenko}, {Ganeshalingam}, {Leonard}, {Li}, \& {Thomas}}]{Poznanski10}
{Poznanski}, D., {Chornock}, R., {Nugent}, P.~E., {et~al.} 2010, Science, 327,
  58, \dodoi{10.1126/science.1181709}

\bibitem[{{Prasad} \& {Chengalur}(2012)}]{Prasad12}
{Prasad}, J., \& {Chengalur}, J. 2012, Experimental Astronomy, 33, 157,
  \dodoi{10.1007/s10686-011-9279-5}

\bibitem[{{Prentice} {et~al.}(2018){Prentice}, {Maguire}, {Smartt}, {Magee},
  {Schady}, {Sim}, {Chen}, {Clark}, {Colin}, {Fulton}, {McBrien}, {O'Neill},
  {Smith}, {Ashall}, {Chambers}, {Denneau}, {Flewelling}, {Heinze}, {Holoien},
  {Huber}, {Kochanek}, {Mazzali}, {Prieto}, {Rest}, {Shappee}, {Stalder},
  {Stanek}, {Stritzinger}, {Thompson}, \& {Tonry}}]{Prentice18}
{Prentice}, S.~J., {Maguire}, K., {Smartt}, S.~J., {et~al.} 2018, \apj, 865,
  L3, \dodoi{10.3847/2041-8213/aadd90}

\bibitem[{{Pursiainen} {et~al.}(2018){Pursiainen}, {Childress}, {Smith},
  {Prajs}, {Sullivan}, {Davis}, {Foley}, {Asorey}, {Calcino}, {Carollo},
  {Curtin}, {D'Andrea}, {Glazebrook}, {Gutierrez}, {Hinton}, {Hoormann},
  {Inserra}, {Kessler}, {King}, {Kuehn}, {Lewis}, {Lidman}, {Macaulay},
  {M{\"o}ller}, {Nichol}, {Sako}, {Sommer}, {Swann}, {Tucker}, {Uddin},
  {Wiseman}, {Zhang}, {Abbott}, {Abdalla}, {Allam}, {Annis}, {Avila}, {Brooks},
  {Buckley-Geer}, {Burke}, {Carnero Rosell}, {Carrasco Kind}, {Carretero},
  {Castander}, {Cunha}, {Davis}, {De Vicente}, {Diehl}, {Doel}, {Eifler},
  {Flaugher}, {Fosalba}, {Frieman}, {Garc{\'\i}a-Bellido}, {Gruen}, {Gruendl},
  {Gutierrez}, {Hartley}, {Hollowood}, {Honscheid}, {James}, {Jeltema},
  {Kuropatkin}, {Li}, {Lima}, {Maia}, {Martini}, {Menanteau}, {Ogando},
  {Plazas}, {Roodman}, {Sanchez}, {Scarpine}, {Schindler}, {Smith},
  {Soares-Santos}, {Sobreira}, {Suchyta}, {Swanson}, {Tarle}, {Tucker}, \&
  {Walker}}]{Pursiainen18}
{Pursiainen}, M., {Childress}, M., {Smith}, M., {et~al.} 2018, \mnras, 481,
  894, \dodoi{10.1093/mnras/sty2309}

\bibitem[{{Quimby} {et~al.}(2013){Quimby}, {Yuan}, {Akerlof}, \&
  {Wheeler}}]{Quimby13}
{Quimby}, R.~M., {Yuan}, F., {Akerlof}, C., \& {Wheeler}, J.~C. 2013, \mnras,
  431, 912, \dodoi{10.1093/mnras/stt213}

\bibitem[{{Rau} {et~al.}(2009){Rau}, {Kulkarni}, {Law}, {Bloom}, {Ciardi},
  {Djorgovski}, {Fox}, {Gal-Yam}, {Grillmair}, {Kasliwal}, {Nugent}, {Ofek},
  {Quimby}, {Reach}, {Shara}, {Bildsten}, {Cenko}, {Drake}, {Filippenko},
  {Helfand}, {Helou}, {Howell}, {Poznanski}, \& {Sullivan}}]{Rau09PTF}
{Rau}, A., {Kulkarni}, S.~R., {Law}, N.~M., {et~al.} 2009, \pasp, 121, 1334,
  \dodoi{10.1086/605911}

\bibitem[{{Readhead}(1994)}]{Readhead94}
{Readhead}, A.~C.~S. 1994, \apj, 426, 51, \dodoi{10.1086/174038}

\bibitem[{{Rees}(1988)}]{Rees88}
{Rees}, M.~J. 1988, \nat, 333, 523, \dodoi{10.1038/333523a0}

\bibitem[{{Reines} \& {Volonteri}(2015)}]{Reines15}
{Reines}, A.~E., \& {Volonteri}, M. 2015, \apj, 813, 82,
  \dodoi{10.1088/0004-637X/813/2/82}

\bibitem[{{Rest} {et~al.}(2018){Rest}, {Garnavich}, {Khatami}, {Kasen},
  {Tucker}, {Shaya}, {Olling}, {Mushotzky}, {Zenteno}, {Margheim},
  {Strampelli}, {James}, {Smith}, {F{\"o}rster}, \& {Villar}}]{Rest18}
{Rest}, A., {Garnavich}, P.~M., {Khatami}, D., {et~al.} 2018, Nature Astronomy,
  2, 307, \dodoi{10.1038/s41550-018-0423-2}

\bibitem[{{Reynolds} {et~al.}(2016){Reynolds}, {Dong}, {Fraser}, {Stritzinger},
  {Mattila}, {Lundqvist}, {Elias-Rosa}, {Harmanen}, {Kangas}, {Somero},
  {Kuncarayakti}, \& {Taddia}}]{Reynolds16}
{Reynolds}, T., {Dong}, S., {Fraser}, M., {et~al.} 2016, The Astronomer's
  Telegram, 9645, 1

\bibitem[{{Rivera Sandoval} {et~al.}(2018){Rivera Sandoval}, {Maccarone},
  {Corsi}, {Brown}, {Pooley}, \& {Wheeler}}]{RiveraSandoval18}
{Rivera Sandoval}, L.~E., {Maccarone}, T.~J., {Corsi}, A., {et~al.} 2018,
  \mnras, 480, L146, \dodoi{10.1093/mnrasl/sly145}

\bibitem[{{Rybicki} \& {Lightman}(1979)}]{Rybicki1979}
{Rybicki}, G.~B., \& {Lightman}, A.~P. 1979, {Radiative processes in
  astrophysics} (J. Wiley \& Sons, Wiley-VCH)

\bibitem[{{Saxton} {et~al.}(2017){Saxton}, {Read}, {Komossa}, {Lira},
  {Alexander}, \& {Wieringa}}]{Saxton17}
{Saxton}, R.~D., {Read}, A.~M., {Komossa}, S., {et~al.} 2017, \aap, 598, A29,
  \dodoi{10.1051/0004-6361/201629015}

\bibitem[{{Schlafly} \& {Finkbeiner}(2011)}]{Schlafly11}
{Schlafly}, E.~F., \& {Finkbeiner}, D.~P. 2011, \apj, 737, 103,
  \dodoi{10.1088/0004-637X/737/2/103}

\bibitem[{{Schulze} {et~al.}(2018){Schulze}, {Kr{\"u}hler}, {Leloudas},
  {Gorosabel}, {Mehner}, {Buchner}, {Kim}, {Ibar}, {Amor{\'\i}n},
  {Herrero-Illana}, {Anderson}, {Bauer}, {Christensen}, {de Pasquale}, {de
  Ugarte Postigo}, {Gallazzi}, {Hjorth}, {Morrell}, {Malesani}, {Sparre},
  {Stalder}, {Stark}, {Th{\"o}ne}, \& {Wheeler}}]{Schulze18}
{Schulze}, S., {Kr{\"u}hler}, T., {Leloudas}, G., {et~al.} 2018, \mnras, 473,
  1258, \dodoi{10.1093/mnras/stx2352}

\bibitem[{{Scott} \& {Readhead}(1977)}]{Scott1977}
{Scott}, M.~A., \& {Readhead}, A.~C.~S. 1977, \mnras, 180, 539,
  \dodoi{10.1093/mnras/180.4.539}

\bibitem[{{Sedov}(1946)}]{Sedov46}
{Sedov}, L. 1946, J. Appl. Math. Mech., 10, 241

\bibitem[{{Shappee} {et~al.}(2014){Shappee}, {Prieto}, {Stanek}, {Kochanek},
  {Holoien}, {Jencson}, {Basu}, {Beacom}, {Szczygiel}, {Pojmanski},
  {Brimacombe}, {Dubberley}, {Elphick}, {Foale}, {Hawkins}, {Mullins},
  {Rosing}, {Ross}, \& {Walker}}]{Shappee14}
{Shappee}, B., {Prieto}, J., {Stanek}, K.~Z., {et~al.} 2014, in American
  Astronomical Society Meeting Abstracts, Vol. 223, American Astronomical
  Society Meeting Abstracts \#223, 236.03

\bibitem[{{Shen} {et~al.}(2010){Shen}, {Kasen}, {Weinberg}, {Bildsten}, \&
  {Scannapieco}}]{Shen10}
{Shen}, K.~J., {Kasen}, D., {Weinberg}, N.~N., {Bildsten}, L., \&
  {Scannapieco}, E. 2010, \apj, 715, 767, \dodoi{10.1088/0004-637X/715/2/767}

\bibitem[{{Shivvers} {et~al.}(2016){Shivvers}, {Zheng}, {Mauerhan}, {Kleiser},
  {Van Dyk}, {Silverman}, {Graham}, {Kelly}, {Filippenko}, \&
  {Kumar}}]{Shivvers16}
{Shivvers}, I., {Zheng}, W.~K., {Mauerhan}, J., {et~al.} 2016, \mnras, 461,
  3057, \dodoi{10.1093/mnras/stw1528}

\bibitem[{{Skrutskie} {et~al.}(2006){Skrutskie}, {Cutri}, {Stiening},
  {Weinberg}, {Schneider}, {Carpenter}, {Beichman}, {Capps}, {Chester},
  {Elias}, {Huchra}, {Liebert}, {Lonsdale}, {Monet}, {Price}, {Seitzer},
  {Jarrett}, {Kirkpatrick}, {Gizis}, {Howard}, {Evans}, {Fowler}, {Fullmer},
  {Hurt}, {Light}, {Kopan}, {Marsh}, {McCallon}, {Tam}, {Van Dyk}, \&
  {Wheelock}}]{Skrutskie06}
{Skrutskie}, M.~F., {Cutri}, R.~M., {Stiening}, R., {et~al.} 2006, \aj, 131,
  1163, \dodoi{10.1086/498708}

\bibitem[{{Slysh}(1990)}]{Slysh90}
{Slysh}, V.~I. 1990, Soviet Astronomy Letters, 16, 339

\bibitem[{{Smith}(2014)}]{Smith14}
{Smith}, N. 2014, \araa, 52, 487, \dodoi{10.1146/annurev-astro-081913-040025}

\bibitem[{{Soderberg} {et~al.}(2010{\natexlab{a}}){Soderberg}, {Brunthaler},
  {Nakar}, {Chevalier}, \& {Bietenholz}}]{Soderberg10b}
{Soderberg}, A.~M., {Brunthaler}, A., {Nakar}, E., {Chevalier}, R.~A., \&
  {Bietenholz}, M.~F. 2010{\natexlab{a}}, \apj, 725, 922,
  \dodoi{10.1088/0004-637X/725/1/922}

\bibitem[{{Soderberg} {et~al.}(2006{\natexlab{a}}){Soderberg}, {Chevalier},
  {Kulkarni}, \& {Frail}}]{Soderberg06d}
{Soderberg}, A.~M., {Chevalier}, R.~A., {Kulkarni}, S.~R., \& {Frail}, D.~A.
  2006{\natexlab{a}}, \apj, 651, 1005, \dodoi{10.1086/507571}

\bibitem[{{Soderberg} {et~al.}(2005){Soderberg}, {Kulkarni}, {Berger},
  {Chevalier}, {Frail}, {Fox}, \& {Walker}}]{Soderberg05}
{Soderberg}, A.~M., {Kulkarni}, S.~R., {Berger}, E., {et~al.} 2005, \apj, 621,
  908, \dodoi{10.1086/427649}

\bibitem[{{Soderberg} {et~al.}(2006{\natexlab{b}}){Soderberg}, {Kulkarni},
  {Nakar}, {Berger}, {Cameron}, {Fox}, {Frail}, {Gal-Yam}, {Sari}, {Cenko},
  {Kasliwal}, {Chevalier}, {Piran}, {Price}, {Schmidt}, {Pooley}, {Moon},
  {Penprase}, {Ofek}, {Rau}, {Gehrels}, {Nousek}, {Burrows}, {Persson}, \&
  {McCarthy}}]{Soderberg06c}
{Soderberg}, A.~M., {Kulkarni}, S.~R., {Nakar}, E., {et~al.}
  2006{\natexlab{b}}, \nat, 442, 1014, \dodoi{10.1038/nature05087}

\bibitem[{{Soderberg} {et~al.}(2010{\natexlab{b}}){Soderberg}, {Chakraborti},
  {Pignata}, {Chevalier}, {Chandra}, {Ray}, {Wieringa}, {Copete}, {Chaplin},
  {Connaughton}, {Barthelmy}, {Bietenholz}, {Chugai}, {Stritzinger}, {Hamuy},
  {Fransson}, {Fox}, {Levesque}, {Grindlay}, {Challis}, {Foley}, {Kirshner},
  {Milne}, \& {Torres}}]{Soderberg10}
{Soderberg}, A.~M., {Chakraborti}, S., {Pignata}, G., {et~al.}
  2010{\natexlab{b}}, \nat, 463, 513, \dodoi{10.1038/nature08714}

\bibitem[{{Strubbe} \& {Quataert}(2009)}]{Strubbe09}
{Strubbe}, L.~E., \& {Quataert}, E. 2009, \mnras, 400, 2070,
  \dodoi{10.1111/j.1365-2966.2009.15599.x}

\bibitem[{{Svensson} {et~al.}(2010){Svensson}, {Levan}, {Tanvir}, {Fruchter},
  \& {Strolger}}]{Svensson10}
{Svensson}, K.~M., {Levan}, A.~J., {Tanvir}, N.~R., {Fruchter}, A.~S., \&
  {Strolger}, L.~G. 2010, \mnras, 405, 57,
  \dodoi{10.1111/j.1365-2966.2010.16442.x}

\bibitem[{{Tampo} {et~al.}(2020){Tampo}, {Tanaka}, {Maeda}, {Yasuda},
  {Tominaga}, {Jiang}, {Moriya}, {Morokuma}, {Suzuki}, {Takahashi}, {Kokubo},
  \& {Kawana}}]{Tampo20}
{Tampo}, Y., {Tanaka}, M., {Maeda}, K., {et~al.} 2020, arXiv e-prints,
  arXiv:2003.02669.
\newblock \doarXiv{2003.02669}

\bibitem[{{Tan} {et~al.}(2001){Tan}, {Matzner}, \& {McKee}}]{Tan01}
{Tan}, J.~C., {Matzner}, C.~D., \& {McKee}, C.~F. 2001, \apj, 551, 946,
  \dodoi{10.1086/320245}

\bibitem[{{Tanaka} {et~al.}(2016){Tanaka}, {Tominaga}, {Morokuma}, {Yasuda},
  {Furusawa}, {Baklanov}, {Blinnikov}, {Moriya}, {Doi}, {Jiang}, {Kato},
  {Kikuchi}, {Kuncarayakti}, {Nagao}, {Nomoto}, \& {Taniguchi}}]{Tanaka16}
{Tanaka}, M., {Tominaga}, N., {Morokuma}, T., {et~al.} 2016, \apj, 819, 5,
  \dodoi{10.3847/0004-637X/819/1/5}

\bibitem[{{Tartaglia} {et~al.}(2018){Tartaglia}, {Sand}, {Valenti}, {Wyatt},
  {Anderson}, {Arcavi}, {Ashall}, {Botticella}, {Cartier}, {Chen}, {Cikota},
  {Coulter}, {Della Valle}, {Foley}, {Gal-Yam}, {Galbany}, {Gall}, {Haislip},
  {Harmanen}, {Hosseinzadeh}, {Howell}, {Hsiao}, {Inserra}, {Jha}, {Kankare},
  {Kilpatrick}, {Kouprianov}, {Kuncarayakti}, {Maccarone}, {Maguire},
  {Mattila}, {Mazzali}, {McCully}, {Melandri}, {Morrell}, {Phillips},
  {Pignata}, {Piro}, {Prentice}, {Reichart}, {Rojas-Bravo}, {Smartt}, {Smith},
  {Sollerman}, {Stritzinger}, {Sullivan}, {Taddia}, \& {Young}}]{Tartaglia18}
{Tartaglia}, L., {Sand}, D.~J., {Valenti}, S., {et~al.} 2018, \apj, 853, 62,
  \dodoi{10.3847/1538-4357/aaa014}

\bibitem[{{Taylor}(1950)}]{Taylor50}
{Taylor}, G. 1950, Proceedings of the Royal Society of London Series A, 201,
  159, \dodoi{10.1098/rspa.1950.0049}

\bibitem[{{van Eerten} {et~al.}(2012){van Eerten}, {van der Horst}, \&
  {MacFadyen}}]{vanEerten2012}
{van Eerten}, H., {van der Horst}, A., \& {MacFadyen}, A. 2012, \apj, 749, 44,
  \dodoi{10.1088/0004-637X/749/1/44}

\bibitem[{{Vink{\'o}} {et~al.}(2015){Vink{\'o}}, {Yuan}, {Quimby}, {Wheeler},
  {Ramirez-Ruiz}, {Guillochon}, {Chatzopoulos}, {Marion}, \&
  {Akerlof}}]{Vinko15}
{Vink{\'o}}, J., {Yuan}, F., {Quimby}, R.~M., {et~al.} 2015, \apj, 798, 12,
  \dodoi{10.1088/0004-637X/798/1/12}

\bibitem[{von Neumann(1941)}]{vonNeumann41}
von Neumann, J. 1941, The point source solution, {U.S. Government} Document
  {AM-9}, National Defense Research Council, Division B, Washington, DC, USA

\bibitem[{{Whitesides} {et~al.}(2017){Whitesides}, {Lunnan}, {Kasliwal},
  {Perley}, {Corsi}, {Cenko}, {Blagorodnova}, {Cao}, {Cook}, {Doran},
  {Frederiks}, {Fremling}, {Hurley}, {Karamehmetoglu}, {Kulkarni}, {Leloudas},
  {Masci}, {Nugent}, {Ritter}, {Rubin}, {Savchenko}, {Sollerman}, {Svinkin},
  {Taddia}, {Vreeswijk}, \& {Wozniak}}]{Whitesides17}
{Whitesides}, L., {Lunnan}, R., {Kasliwal}, M.~M., {et~al.} 2017, \apj, 851,
  107, \dodoi{10.3847/1538-4357/aa99de}

\bibitem[{{Yang} {et~al.}(2017){Yang}, {Valenti}, {Cappellaro}, {Sand},
  {Tartaglia}, {Corsi}, {Reichart}, {Haislip}, \& {Kouprianov}}]{Yang17}
{Yang}, S., {Valenti}, S., {Cappellaro}, E., {et~al.} 2017, \apjl, 851, L48,
  \dodoi{10.3847/2041-8213/aaa07d}

\bibitem[{{Zauderer} {et~al.}(2013){Zauderer}, {Berger}, {Margutti}, {Pooley},
  {Sari}, {Soderberg}, {Brunthaler}, \& {Bietenholz}}]{Zauderer13}
{Zauderer}, B.~A., {Berger}, E., {Margutti}, R., {et~al.} 2013, \apj, 767, 152,
  \dodoi{10.1088/0004-637X/767/2/152}

\bibitem[{{Zauderer} {et~al.}(2011){Zauderer}, {Berger}, {Soderberg}, {Loeb},
  {Narayan}, {Frail}, {Petitpas}, {Brunthaler}, {Chornock}, {Carpenter},
  {Pooley}, {Mooley}, {Kulkarni}, {Margutti}, {Fox}, {Nakar}, {Patel},
  {Volgenau}, {Culverhouse}, {Bietenholz}, {Rupen}, {Max-Moerbeck}, {Readhead},
  {Richards}, {Shepherd}, {Storm}, \& {Hull}}]{Zauderer11}
{Zauderer}, B.~A., {Berger}, E., {Soderberg}, A.~M., {et~al.} 2011, \nat, 476,
  425, \dodoi{10.1038/nature10366}

\end{thebibliography}

%%%%%%%%%%%%%%%%%%%%%%%%%%%%%%%%%%%%%%%%%%%%%%%%%%%%%%%%%
\appendix 

\section{Observations}

\begin{table}[h]
\centering
    \caption{Radio Observations of \sn{}}
    \label{Tab:radio}
    \begin{tabular}{cccccc}
    \hline
  Start Date & Time$^{\rm{a}}$ & Frequency$^{\rm{b}}$ & Bandwidth & Flux Density$^{\rm{c}}$ & Telescope \\
(UT) & (days) & (GHz) & (GHz) & (mJy) & \\
    \hline
2016-12-14 & 69 & 6.10 & 2.048 & $4.5\pm0.2$ & VLA\\
2017-01-13 & 99 & 1.50 & 1.024 & $1.5\pm0.1$ & VLA\\
2017-01-13 & 99 & 3.00 & 2.048 & $4.3\pm0.2$ & VLA\\
2017-01-13 & 99 & 6.10 & 2.048 & $6.1\pm0.3$ & VLA\\
2017-01-13 & 99 & 9.87 & 4.096 & $4.2\pm0.2$ & VLA\\
2017-03-17 & 162 & 1.50 & 1.024 & $4.7\pm0.6$$^{\rm{d}}$ & VLA\\
2017-03-17 & 162 & 2.94 & 2.048 & $2.9\pm0.2$ & VLA\\
2017-03-17 & 162 & 6.10 & 2.048 & $2.3\pm0.1$ & VLA\\
2017-03-17 & 162 & 9.74 & 4.096 & $1.74\pm0.09$ & VLA\\
2017-03-17 & 162 & 22.00 & 8.192 & $0.56\pm0.03$ & VLA\\
2017-09-14 & 343 & 1.39 & 0.032 & $0.38\pm0.05$ & GMRT\\
2017-09-19 & 348 & 0.33 & 0.032 & $\leq0.375$ & GMRT\\
2017-09-21 & 350 & 0.61 & 0.032 & $0.79\pm0.09$ & GMRT\\
2017-09-28 & 357 & 1.50 & 1.024 & $0.27\pm0.07$ & VLA\\
2017-09-28 & 357 & 3.00 & 2.048 & $0.17\pm0.03$ & VLA\\
2017-09-28 & 357 & 6.05 & 2.048 & $0.07\pm0.01$ & VLA\\
2017-09-28 & 357 & 10.00 & 4.096 & $0.032\pm0.008$ & VLA\\
2018-03-21 & 531 & 1.50 & 1.024 & $\leq0.065$ & VLA\\
2018-03-21 & 531 & 10.00 & 4.096 & $\leq0.018$ & VLA\\
    \hline
    \end{tabular}
\tablecomments{$^{\rm{a}}$ Days since JD 2457668. $^{\rm{b}}$ The table containing flux densities for each of the sub-bands as displayed in Figure \ref{Fig:SED_fits_SSA} is available online. $^{\rm{c}}$ Uncertainties are quoted at 1$\sigma$, and upper-limits are quoted at $3\sigma$. The errors take a systematic uncertainty of 5\% (VLA) or 15\% (GMRT) into account. $^{\rm{d}}$ There was significant RFI in this band.}
\end{table} 

\end{document}